\documentclass[12pt]{iopart}  

\usepackage{graphicx} \usepackage{bm}

\begin{document}

\title[Continuum variational and diffusion quantum Monte Carlo
calculations]{Continuum variational and diffusion quantum Monte Carlo
calculations}

\author{R J Needs, M D Towler, N D Drummond and P L\'opez R\'ios}

\address{Theory of Condensed Matter Group, Cavendish Laboratory, Cambridge CB3
0HE, UK}

\begin{abstract}
This topical review describes the methodology of continuum variational and
diffusion quantum Monte Carlo calculations.  These stochastic methods are
based on many-body wave functions and are capable of achieving very high
accuracy.  The algorithms are intrinsically parallel and well-suited to
petascale computers, and the computational cost scales as a polynomial of the
number of particles.  A guide to the systems and topics which have been
investigated using these methods is given. The bulk of the article is devoted
to an overview of the basic quantum Monte Carlo methods, the forms and
optimisation of wave functions, performing calculations within periodic
boundary conditions, using pseudopotentials, excited-state calculations,
sources of calculational inaccuracy, and calculating energy differences and
forces.
\end{abstract}


\submitto{\JPCM}

\maketitle

\section{Introduction \label{sec:introduction}}

The variational Monte Carlo (VMC) and diffusion Monte Carlo (DMC) methods are
stochastic approaches for evaluating quantum mechanical expectation values
with many-body Hamiltonians and wave functions \cite{foulkes_2001}. VMC and
DMC methods are used for both continuum and lattice systems, but here we
describe their application only to continuum systems.  The main attraction of
these methods is that the computational cost scales as some reasonable power
(normally from the second to fourth power) of the number of particles
\cite{note_exp_scaling}.
This scaling makes it possible to deal with hundreds or even thousands of
particles, allowing applications to condensed matter.

Continuum quantum Monte Carlo (QMC) methods, such as VMC and DMC, occupy a
special place in the hierarchy of computational approaches for modelling
materials.  QMC computations are expensive, which limits their applicability
at present, but they are the most accurate methods known for computing the
energies of large assemblies of interacting quantum particles.  There are many
problems for which the high accuracy achievable with QMC is necessary to give
a faithful description of the underlying science.  Most of our work is
concerned with correlated electron systems, but these methods can be applied
to any combination of fermion and boson particles with any inter-particle
potentials and external fields \textit{etc.}  Being based on many-body wave
functions, these are zero-temperature methods, and for finite temperatures one
must use other approaches such as those based on density matrices.

Both the VMC and DMC methods are variational, so that the calculated energy is
above the true ground state energy.  The computational costs of VMC and DMC
calculations scale similarly with the number of particles studied, but the
prefactor is larger for the more accurate DMC method.  QMC algorithms are
intrinsically parallel and are ideal candidates for taking advantage of the
petascale computers (10$^{15}$ flops) which are becoming available now and the
exascale computers (10$^{18}$ flops) which will be available one day.

DMC has been applied to a wide variety of continuum systems.  A partial list
of topics investigated within DMC and some references to milestone papers are
given below.
\begin{itemize}
\item Three-dimensional electron gas
\cite{ceperley_1980,moroni_1995,zong_2002,drummond_2004}.
\item Two-dimensional electron gas
\cite{tanatar_1989,moroni_1992,attaccalite_2002,drummond_2008_2d}.
\item The equation of state and other properties of liquid $^3$He
\cite{lee_1981,holzmann_2006}.
\item Structure of nuclei \cite{carlson_2007}.
\item Pairing in ultra-cold atomic gases
\cite{carlson_2003,astrakharchik_2004,carlson_2008}.
\item Reconstruction of a crystalline surface \cite{healy_2001} and molecules
on surfaces \cite{filippi_2002,kim_2006}.
\item Quantum dots \cite{ghosal_2006}.
\item Band structures of insulators
\cite{mitas_1994,williamson_1998,towler_2000}.
\item Transition metal oxide chemistry
\cite{towler_2003,wagner_2003,wagner_2007a}.
\item Optical band gaps of nanocrystals
\cite{williamson_2002,drummond_2005_dia}.
\item Defects in semiconductors \cite{leung_1999,hood_2003,alfe_2005a}.
\item Solid state structural phase transitions \cite{alfe_2005b}.
\item Equations of state of solids
\cite{natoli_1993,delaney_2006,maezono_2007a,pozzo_2008}.
\item Binding of molecules and their excitation energies
\cite{manten_2001,grossman_2002,aspuru-guzik_2004,gurtubay_2006,gurtubay_2007}.
\item Studies of exchange-correlation
\cite{hood_1997,hood_1998,nekovee_2001,nekovee_2003}.
\end{itemize}


The same basic QMC algorithm can be used for each of the applications
mentioned above with only minor modifications.  The complexity and
sophistication of the computer codes arises not from the algorithm itself,
which is in fact quite simple, but from the diversity of the Hamiltonians and
many-body wave functions which are involved.  A number of computer codes are
currently available for performing continuum QMC calculations of the type
described here \cite{qmc_wiki}.  We have developed the \textsc{casino} code
\cite{casino}, which can deal with systems of different dimensionalities,
various interactions including the Coulomb potential,
external fields, mixtures of particles of different types and different types
of many-body wave function.

The VMC and DMC methods are described in section \ref{sec:qmc} and
the types of many-body wave function we use are described in section
\ref{sec:psi_trial}.  The optimisation of parameters in wave functions
using stochastic methods which are both subtle and unique to the field
is described in section \ref{sec:optimise_psi_trial}.  QMC
calculations within periodic boundary conditions are described in
section \ref{sec:pbc}, the use of pseudopotentials in QMC calculations
is discussed in section \ref{sec:pseudopots} and excited-state DMC
calculations are briefly described in section
\ref{sec:excited_states}.  The scaling of the QMC methods with system
size is discussed in section \ref{sec:scaling}.  Sources of errors in
the DMC method and practical methods for handling errors in QMC
results are described in section \ref{subsec:errors}.  In section
\ref{sec:other expectation values} we describe how to evaluate other
expectation values apart from the energy.  Section \ref{sec:energy
differences and energy derivatives} deals with the calculation of
energy differences and energy derivatives in the VMC and DMC methods,
and we make our final remarks in section \ref{sec:conclusions}.

\section{Quantum Monte Carlo methods}
\label{sec:qmc}

The VMC method is conceptually very simple. The energy is calculated as the
expectation value of the Hamiltonian with an approximate many-body trial wave
function.  In the more sophisticated DMC method the estimate of the ground
state energy is improved by performing a process described by the evolution of
the wave function in imaginary time.  Throughout this article we will consider
only systems with spin-independent Hamiltonians and collinear spins.  We will
also restrict the discussion to systems with time-reversal symmetry, for which
the wave function may be chosen to be real.  It is, however, straightforward
to generalise the VMC algorithm to work with complex wave functions, and only
a little more complicated to generalise the DMC algorithm to work with them
\cite{ortiz_1993}.

\subsection{The VMC method}
\label{subsec:vmc}

The variational theorem of quantum mechanics states that, for a real, proper
\cite{footnote:proper} trial wave function $\Psi_{\rm T}$, the variational
energy,
\begin{eqnarray}
\label{eq:variational_energy}
E_{\rm V} = \frac{\int \Psi_{\rm T}({\bf R}) \hat{H} \Psi_{\rm T}({\bf R})\,
d{\bf R}}{\int \Psi_{\rm T}^2({\bf R}) \, d{\bf R}} \;,
\end{eqnarray}
is an upper bound on the exact ground state energy $E_0$, \textit{i.e.},
$E_{\rm V} \geq E_0$.  In equation (\ref{eq:variational_energy}), $\hat{H}$ is
the many-body Hamiltonian and ${\bf R}$ denotes a $3N$-dimensional vector of
particle coordinates.  As discussed in section \ref{subsec:Slater-Jastrow wave
functions}, the spin variables in equation (\ref{eq:variational_energy}) are
implicitly summed over.

To facilitate the stochastic evaluation, $E_{\rm V}$ is written as
\begin{eqnarray}
\label{eq:variational_energy_2}
E_{\rm V} = \int p({\bf R}) E_{\rm L}({\bf R}) \, d{{\bf R}} \;,
\end{eqnarray}
where the probability distribution $p$ is
\begin{eqnarray}
\label{eq:distribution}
p({\bf R}) = \frac{\Psi_{\rm T}^2({\bf R})}{\int \Psi_{\rm T}^2({\bf
R}^{\prime}) \, d{\bf R}^{\prime}} \;,
\end{eqnarray}
and the local energy,
\begin{eqnarray}
\label{eq:local_energy}
E_{\rm L}({\bf R}) = \Psi_{\rm T}^{-1} \hat{H} \Psi_{\rm T} \;.
\end{eqnarray}
is straightforward to evaluate at any ${\bf R}$.

In VMC the Metropolis algorithm \cite{metropolis_1953} is used to sample the
probability distribution $p({\bf R})$. Let the electron configuration at a
particular step be ${\bf R}^\prime$.  A new configuration ${\bf R}$ is drawn
from the probability density $T({\bf R} \leftarrow {\bf R}^\prime)$, and the
move is accepted with probability
\begin{equation} A({\bf R}\leftarrow{\bf R}^\prime)= \min \left\{
1,\frac{T({\bf R}^\prime \leftarrow {\bf R})\Psi_{\rm T}^2({\bf R})}{T({\bf R}
\leftarrow {\bf R}^\prime)\Psi_{\rm T}^2({\bf R}^\prime)} \right\}.
\end{equation}
It can easily be verified that this algorithm satisfies the \textit{detailed
balance} condition
\begin{equation}
\Psi_{\rm T}^2({\bf R})T({\bf R}^\prime \leftarrow {\bf R}) A({\bf R}^\prime
\leftarrow {\bf R}) = \Psi_{\rm T}^2({\bf R}^\prime)T({\bf R} \leftarrow {\bf
R}^\prime) A({\bf R} \leftarrow {\bf R}^\prime).
\end{equation}
Hence $p({\bf R})$ is the equilibrium configuration distribution of this
Markov process and, so long as the transition probability is ergodic (i.e., it
is possible to reach any point in configuration space in a finite number of
moves), it can be shown that the process will converge to this equilibrium
distribution.  Once equilibrium has been reached, the configurations are
distributed as $p({\bf R})$, but successive configurations along the random
walk are in general correlated.

The variational energy is estimated as
\begin{eqnarray}
\label{eq:variational_energy_3}
E_{\rm V} \simeq \frac{1}{M} \sum_{i=1}^M E_{\rm L}({\bf R}_i),
\end{eqnarray}
where $M$ configurations ${\bf R}_i$ have been generated after equilibration.
The serial correlation of the configurations and therefore local energies
$E_{\rm L}({\bf R}_i)$ complicates the calculation of the statistical error on
the energy estimate: see section \ref{subsec:statistical errors}.  Other
expectation values may be evaluated in a similar manner to the energy.

Equation (\ref{eq:variational_energy_2}) is an importance sampling
transformation of equation (\ref{eq:variational_energy}).  Equation
(\ref{eq:variational_energy_2}) exhibits the zero variance property: as the
trial wave function approaches an exact eigenfunction ($\Psi_{\rm T}
\rightarrow \phi_i$), the local energy approaches the corresponding
eigenenergy, $E_i$, everywhere in configuration space.  As $\Psi_{\rm T}$ is
improved, $E_{\rm L}$ becomes a smoother function of ${\bf R}$ and the number
of sampling points, $M$, required to achieve an accurate estimate of $E_{\rm
V}$ is reduced.

VMC is a simple and elegant method.  There are no restrictions on the form of
trial wave function which can be used and it does not suffer from a fermion
sign problem.  However, even if the underlying physics is well understood it
is often difficult to prepare trial wave functions of equivalent accuracy for
two different systems, and therefore the VMC estimate of the energy difference
between them will be biased.  We use the VMC method mostly to optimise
parameters in trial wave functions (see section \ref{sec:optimise_psi_trial})
and our main calculations are performed with the more sophisticated DMC
method, which is described in the next section.

\subsection{The DMC method}
\label{subsec:dmc}

In DMC the operator $\exp(-t \hat{H})$ is used to project out the ground state
from the initial state.  This can be viewed as solving the imaginary-time
Schr\"odinger equation, which for electrons is
\begin{eqnarray}
\label{eq:imaginary_time_se}
-\frac{\partial}{\partial t} \Phi({\bf R},t) = \left(\hat{H} - E_{\rm T}
    \right) \Phi({\bf R},t) = \left(-\frac{1}{2} \nabla^2_{\bf R}+ V({\bf R})
    - E_{\rm T} \right) \Phi({\bf R},t) \;,
\end{eqnarray}
where $t$ is a real variable measuring the progress in imaginary time, $V$ is
the potential energy (assumed to be local for the time being), and $E_{\rm T}$
is an arbitrary energy offset known as the reference energy.  Throughout this
article we use Hartree atomic units where $m_e = \hbar = |e| = 4\pi \epsilon_0
= 1$, where $m_e$ is the mass of the electron and $e$ is its charge.  Equation
(\ref{eq:imaginary_time_se}) can be solved formally by expanding $\Phi({\bf
R},t)$ in the eigenstates $\phi_i$ of the Hamiltonian,
\begin{eqnarray}
\label{eq:expansion}
\Phi({\bf R},t) = \sum_i c_i(t) \phi_i({\bf R}) \;,
\end{eqnarray}
which leads to
\begin{eqnarray}
\label{eq:expansion_2}
\Phi({\bf R},t) = \sum_i \exp[-(E_i-E_{\rm T})t] \, c_i(0) \phi_i({\bf R}) \;.
\end{eqnarray}
For long times one finds
\begin{eqnarray}
\label{eq:expansion_3}
\Phi({\bf R},t\rightarrow \infty) \simeq \exp[-(E_0-E_{\rm T})t] \, c_0(0)
\phi_0({\bf R}) \;,
\end{eqnarray}
which is proportional to the ground state wave function, $\phi_0$.

The Hamiltonian is the sum of kinetic and potential terms: $\hat{H} = -(1/2)
\nabla^2_{\bf R}+ V({\bf R})$.  Suppose for a moment that we can interpret the
initial state, $\sum_i c_i(0)\phi_i$, as a probability distribution.  If we
neglect the potential term then the imaginary-time Schr\"odinger equation
(\ref{eq:imaginary_time_se}) reduces to a diffusion equation in the
configuration space. If, on the other hand, we neglect the kinetic term,
(\ref{eq:imaginary_time_se}) reduces to a rate equation.  It should not be
surprising that a short time slice of the imaginary-time evolution can be
simulated by taking a population of configurations $\{{\bf R}_i\}$ and
subjecting them to random hops to simulate the diffusion process, and
``birth'' and ``death'' of configurations to simulate the rate process.  By
``birth'' and ``death'' we mean replicating some configurations and deleting
others at the appropriate rates, a process which is often referred to as
``branching''.

Unfortunately the wave function cannot in general be interpreted as a
probability distribution.  A wave function for two or more identical fermions
must have positive and negative regions, as should an excited state of any
system.  One can construct algorithms which are formally exact using two
distributions of configurations with positive and negative weights
\cite{kalos_2005}, but they are inefficient and the scaling of the
computational cost with system size is unclear.

The fixed-node approximation \cite{anderson_1975,anderson_1976} provides a way
to evade the sign problem.  (In a 3D system, the nodal surface is the
$(3N-1)$-dimensional surface on which the wave function is zero and across
which it changes sign.)  The fixed-node approximation is equivalent to placing
an infinite repulsive potential barrier on the nodal surface of the trial wave
function which is sufficiently strong to force the wave function to be zero on
the nodal surface.  In effect we solve the Schr\"odinger equation exactly
within each pocket enclosed by the nodal surface, subject to the boundary
condition that the wave function is zero on the nodal surface.  The infinite
repulsive potential barrier has no effect if the trial nodal surface is placed
correctly but, if it is not, the energy is always raised.  It follows that the
DMC energy is always less than or equal to the VMC energy with the same trial
wave function, and always greater than or equal to the exact ground-state
energy.


The fixed-node DMC algorithm described above is extremely inefficient and a
vastly superior algorithm can be obtained by introducing an importance
sampling transformation \cite{grimm_1971,kalos_1974}. Consider the mixed
distribution,
\begin{eqnarray}
\label{eq:f}
f({\bf R},t) = \Psi_{\rm T}({\bf R}) \Phi({\bf R},t) \;,
\end{eqnarray}
which has the same sign everywhere if and only if the nodal surface of
$\Phi({\bf R},t)$ equals that of $\Psi_{\rm T}({\bf R})$.  Substituting in
equation (\ref{eq:imaginary_time_se}) for $\Phi$ we obtain
\begin{eqnarray}
\label{eq:importance_sampled_imaginary_time_se}
-\frac{\partial f}{\partial t} = -\frac{1}{2} \nabla_{\bf R}^2 f + \nabla_{\bf
 R} \cdot [{\bf v}f] + [E_{\rm L}-E_{\rm T}]f \;,
\end{eqnarray}
where the $3N$-dimensional drift velocity is defined as
\begin{eqnarray}
\label{eq:drift_velocity}
{\bf v}({\bf R}) = \Psi_{\rm T}^{-1}({\bf R}) \nabla_{\bf R} \Psi_{\rm T}({\bf
R}) \;.
\end{eqnarray}
The three terms on the right-hand side of equation
(\ref{eq:importance_sampled_imaginary_time_se}) correspond to diffusion,
drift and branching processes, respectively.  The importance sampling
transformation has several consequences.  First, the density of configurations
is increased where $|\Psi_{\rm T}|$ is large, so that the more important parts
of the wave function are sampled more often.  Second, the rate of branching is
now controlled by the local energy which is normally a much smoother function
than the potential energy.  This is particularly important for the Coulomb
interaction, which diverges when particles are coincident.  The importance
sampling transformation, together with an algorithm that imposes $f({\bf R},t)
\ge 0$, ensures that $\Psi_{\rm T}$ and $\Phi({\bf R},t)$ have the same nodal
surfaces, as can be seen in equation (\ref{eq:f}).  The importance sampling
transformation also reduces the statistical error bar on the estimate of the
energy and leads to a zero variance property analogous to that in VMC\@.

The importance-sampled imaginary-time Schr\"odinger equation may be written in
integral form:
\begin{eqnarray}
\label{eq:time evolution of f}
f({\bf R},t) = \int G({\bf R} \leftarrow {\bf R}^{\prime}, t-t^{\prime})
f({\bf R}^{\prime},t^{\prime}) \, d{\bf R}^{\prime} \;,
\end{eqnarray}
where the Green's function $G({\bf R} \leftarrow {\bf R}^{\prime},
t-t^{\prime})$ is a solution of equation
(\ref{eq:importance_sampled_imaginary_time_se}) satisfying the initial
condition $G({\bf R} \leftarrow {\bf R}^{\prime}, 0)=\delta({\bf R} - {\bf
R}^{\prime})$.  The exact Green's function can be sampled using the Green's
function Monte Carlo (GFMC) algorithm developed by Kalos and coworkers
\cite{kalos_1962,kalos_1967,kalos_1974,ceperley_1986b,schmidt_1987}.

Let us interpret $f({\bf R},t)$ as the probability distribution of a discrete
population of $P$ configurations with positive weights:
\begin{equation}
\label{eq:f_discrete}
f({\bf R},t) = \left< \sum_{p=1}^P w_p(t) \, \delta[{\bf R}-{\bf R}_p(t)]
\right>,
\end{equation}
where the $p$th configuration at time $t$ has position ${\bf R}_p(t)$ in
configuration space and weight $w_p(t)$, and the angled brackets denote an
ensemble average.  Using equation (\ref{eq:time evolution of f}), the
evolution of $f({\bf R},t)$ to time $t+\tau$ yields
\begin{eqnarray}
\label{eq:f_prime}
f({\bf R},t+\tau) & = & \left< \sum_{p=1}^P w_p(t) \, G[{\bf R} \leftarrow
{\bf R}_p(t),\tau] \right> \nonumber \\ & = & \left< \sum_{p=1}^P w_p(t+\tau)
\, \delta[{\bf R}-{\bf R}_p(t+\tau)] \right>. \end{eqnarray} The dynamics of
the configurations and their weights is governed by the Green's function.

The GFMC algorithm is computationally expensive, but considerably faster
calculations can be made using an approximate Green's functions which becomes
exact in the limit of infinitely small time steps.  Within the short-time
approximation
\begin{eqnarray}
\label{eq:short time G}
G({\bf R} \leftarrow {\bf R}^{\prime}, \tau) \simeq G_{\rm st}({\bf R}
\leftarrow {\bf R}^{\prime}, \tau) = G_{\rm D}({\bf R} \leftarrow {\bf
R}^{\prime}, \tau)G_{\rm B}({\bf R} \leftarrow {\bf R}^{\prime}, \tau) \;,
\end{eqnarray}
where
\begin{eqnarray}
\label{eq:short time GD}
G_{\rm D}({\bf R} \leftarrow {\bf R}^{\prime}, \tau) =
\frac{1}{(2\pi\tau)^{3N/2}} \exp \left(- \frac{\left[ {\bf R}-{\bf R}^{\prime}
- \tau {\bf v}({\bf R}^\prime) \right]^2}{2\tau} \right)
\end{eqnarray}
is the drift-diffusion Green's function and
\begin{eqnarray}
\label{eq:short time GB}
G_{\rm B}({\bf R} \leftarrow {\bf R}^{\prime}, \tau) = \exp \left(
-\frac{\tau}{2} \left[E_{\rm L}({\bf R}) + E_{\rm L}({\bf R}^{\prime}) -
  2E_{\rm T} \right] \right)
\end{eqnarray}
is the branching factor.

The process described by $G_{\rm D}({\bf R}\leftarrow {\bf R}^\prime,\tau)$ is
simulated by making each configuration ${\bf R}^\prime$ in the population
drift through a distance $\tau{\bf v}({\bf R}^\prime)$, then diffuse by a
random distance drawn from a Gaussian distribution of variance $\tau$.  Each
configuration is then copied or deleted in such a fashion that, on average,
$G_{\rm B}({\bf R}\leftarrow {\bf R}^\prime,\tau)$ configurations continue
from the new position ${\bf R}$.  When using the short time approximation,
configurations occasionally attempt to cross the nodal surface but such moves
may simply be rejected.  The short time approximation leads to a dependence of
DMC results on the time step.  It is important to investigate the size of the
time step dependence, and it is common practice to extrapolate the energy to
zero time step: see figure \ref{fig:time_step_errors}.  It turns out that
$G_{\rm st}$ does not precisely satisfy the detailed-balance condition, but it
is standard practice to reinstate detailed balance by incorporating an
accept-reject step.  The importance-sampled fixed-node fermion DMC algorithm
was first used by Ceperley and Alder in their ground-breaking study of the
homogeneous electron gas (HEG) \cite{ceperley_1980}.

\begin{figure}[t]
\centering \includegraphics*[width=1.0\textwidth]{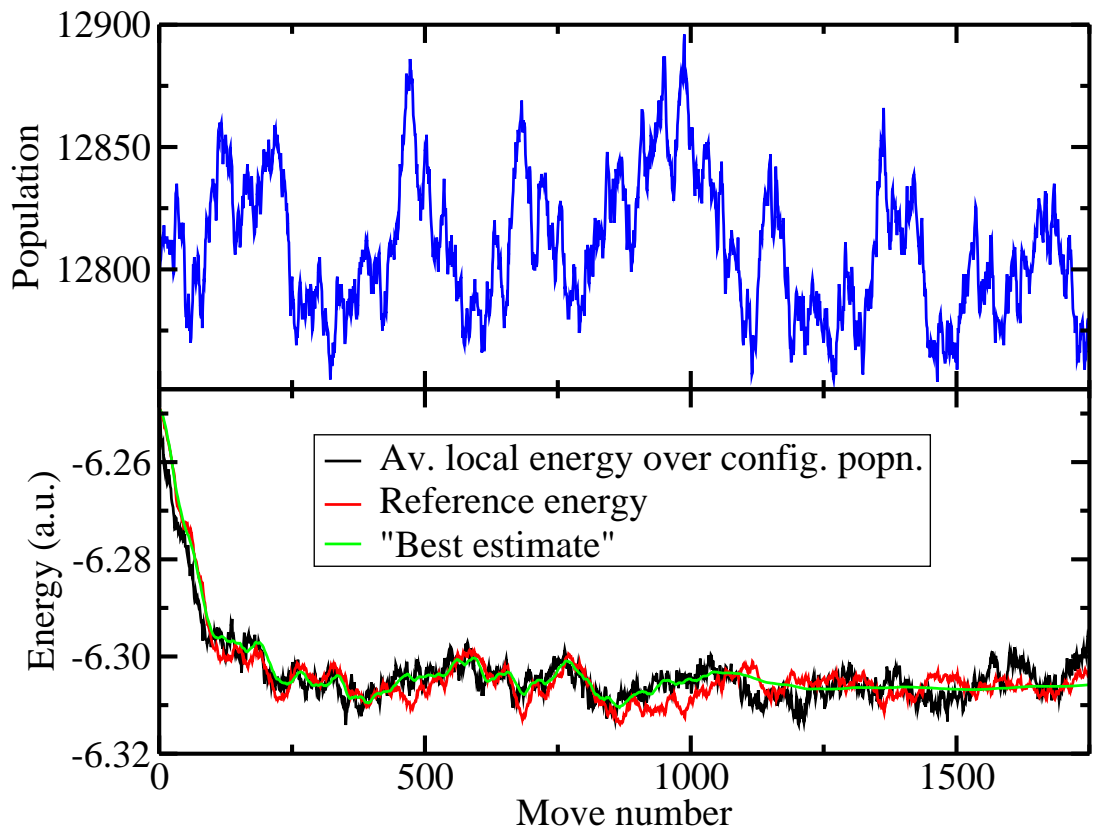}
\caption[]{DMC data for a silane (SiH$_4$) molecule, with the ions represented
by pseudopotentials. The upper panel shows the fluctuations in the population
of configurations arising from the branching process used to simulate equation
(\ref{eq:short time GB}).  The reference energy, $E_{\rm T}$, is altered
during the run to control the population.  Specifically, the reference energy
is set to return the population to the target population (128,000
configurations) on the same time-scale as the autocorrelation period of the
energy data \cite{foulkes_2001}.  The total energy is shown in the lower panel
as a function of the move number.  The black line shows the instantaneous
value of the local energy averaged over the current population of
configurations, the red line is the reference energy $E_{\rm T}$ and the
green line is the best estimate of the DMC energy as the simulation
progresses.  The configurations at move number zero are from the output of a
VMC simulation, and the energy decays rapidly from its initial VMC value of
about -6.250 a.u.\ and reaches a plateau with a DMC energy of about -6.305
a.u.  The data up to move 1000 are deemed to form the equilibration phase, and
are discarded. }
\label{fig:graphit_silane}
\end{figure}

It can be seen that the reference energy $E_{\rm T}$ appears in the branching
factor of equation (\ref{eq:short time GB}).  By adjusting the reference
energy during the simulation we may keep the total population close to a
target value, preventing the population from either increasing exponentially
or dying out.  An example of the behaviour of the total population and the
reference energy can be seen in figure \ref{fig:graphit_silane}
\cite{foulkes_2001}.

Another important aspect of practical implementations is that the particles
are normally moved one at a time in both VMC and DMC algorithms.  The trial
wave function can usually be evaluated more rapidly when a single particle has
been moved than if all particles have been moved, and a longer time step can
be employed for an equivalent time-step error.  The correlation length of the
local energy is shorter for single-particle moves and overall the efficiency
is considerably increased \cite{lopez-rios_2006}.

The initial configurations are normally taken from a VMC calculation and
equilibrated within DMC for a period of imaginary time.  The
importance-sampled DMC algorithm generates configurations asymptotically
distributed according to $f({\bf R})= \Psi_{\rm T}({\bf R})\phi_0({\bf R})$,
where $\phi_0$ is the ground state of the Schr\"odinger equation subject to
the fixed-node boundary condition.  Noting that $\hat{H} \phi_0 = E_0 \phi_0$
everywhere (except on the nodal surface where $\phi_0=0$) the fixed-node DMC
energy can be evaluated using the formula
\begin{eqnarray}
\label{eq:diffusion_energy}
E_{\rm D} \equiv E_0 = \frac{\langle \phi_0 | \hat{H} | \Psi_{\rm T}
\rangle}{\langle \phi_0 | \Psi_{\rm T} \rangle} & = & \frac{\int f({\bf R})
E_{\rm L}({\bf R}) \, d{\bf R}}{\int f({\bf R}) \, d{\bf R}} \\ & \simeq &
\frac{1}{M} \sum_{i=1}^M E_{\rm L}({\bf R}_i) \;.
\end{eqnarray}
Some example DMC data are shown in figure \ref{fig:graphit_silane}.

\section{Trial wave functions}
\label{sec:psi_trial}

Trial wave functions are of central importance in VMC and DMC calculations
because they introduce importance sampling and control both the statistical
efficiency and accuracy obtained.  The accuracy of a DMC calculation depends
on the nodal surface of the trial wave function via the fixed-node
approximation, while in VMC the accuracy depends on the entire trial wave
function.  VMC energies are therefore more sensitive to the quality of the
trial wave function than DMC energies.

\subsection{Slater-Jastrow wave functions}
\label{subsec:Slater-Jastrow wave functions}

QMC calculations require a compact trial wave function which can be evaluated
rapidly.  Most studies of electronic systems have used the Slater-Jastrow
form, in which a pair of up- and down-spin determinants is multiplied by a
Jastrow correlation factor,
\begin{equation}
\label{eq:slater-jastrow}
\Psi_{\rm SJ}({\bf R})\! \! = e^{J({\bf R})}  \det{\left[ \psi_n({\bf
r}_i^{\uparrow})\right]} \det{\left[ \psi_n({\bf r}_j^{\downarrow})\right]} \;,
\end{equation}
where $e^{J}$ is the Jastrow factor and $\det{\left[ \psi_n({\bf
r}_i^{\uparrow}) \right]}$ is a determinant of single-particle orbitals for
the up-spin electrons.  The quality of the single-particle orbitals is very
important, and they are often obtained from density functional theory (DFT) or
Hartree-Fock (HF) calculations.  Note that the spin variables themselves do
not appear in equation (\ref{eq:slater-jastrow}).  Formally the sum over spin
variables in the expectation values in equations (\ref{eq:variational_energy})
and (\ref{eq:diffusion_energy}) has already been performed and the single
determinant with spin variables is replaced by two determinants of up- and
down-spin orbitals whose arguments are the up- and down-spin electron
coordinates ${\bf R}_{\uparrow}$ and ${\bf R}_{\downarrow}$, respectively.
This is explained in more detail in reference \cite{foulkes_2001}.

The Jastrow factor is taken to be symmetric under the interchange of identical
particles and its positivity means that it does not alter the nodal surface of
the trial wave function. The Jastrow factor introduces correlation by making
the wave function depend explicitly on the particle separations.  The optimal
Jastrow factor is normally small when particles with repulsive interactions
(for example, two electrons) are close to one another and large when particles
with attractive interactions (for example, an electron and a positron) are
close to one another.

The Jastrow factor can also be used to ensure that the trial wave function
obeys the Kato cusp conditions \cite{kato_1957}, which leads to smoother
behaviour in the local energy $E_{\rm L}({\bf R})$. When two particles
interacting via the Coulomb potential approach one another, the potential
energy diverges, and therefore the exact wave function $\Psi$ must have a cusp
so that the local kinetic energy $-(1/2) \Psi^{-1}\nabla^2 \Psi$ supplies an
equal and opposite divergence.  It seems very reasonable to enforce the cusp
conditions on trial wave functions because they are obeyed by the exact wave
function.  Imposition of the cusp conditions is in fact very important in both
VMC and DMC calculations because divergences in the local energy lead to poor
statistical behaviour and even instabilities in DMC calculations due to
divergences in the branching factor.

\begin{figure}[t]
\centering \includegraphics*[width=.7\textwidth]{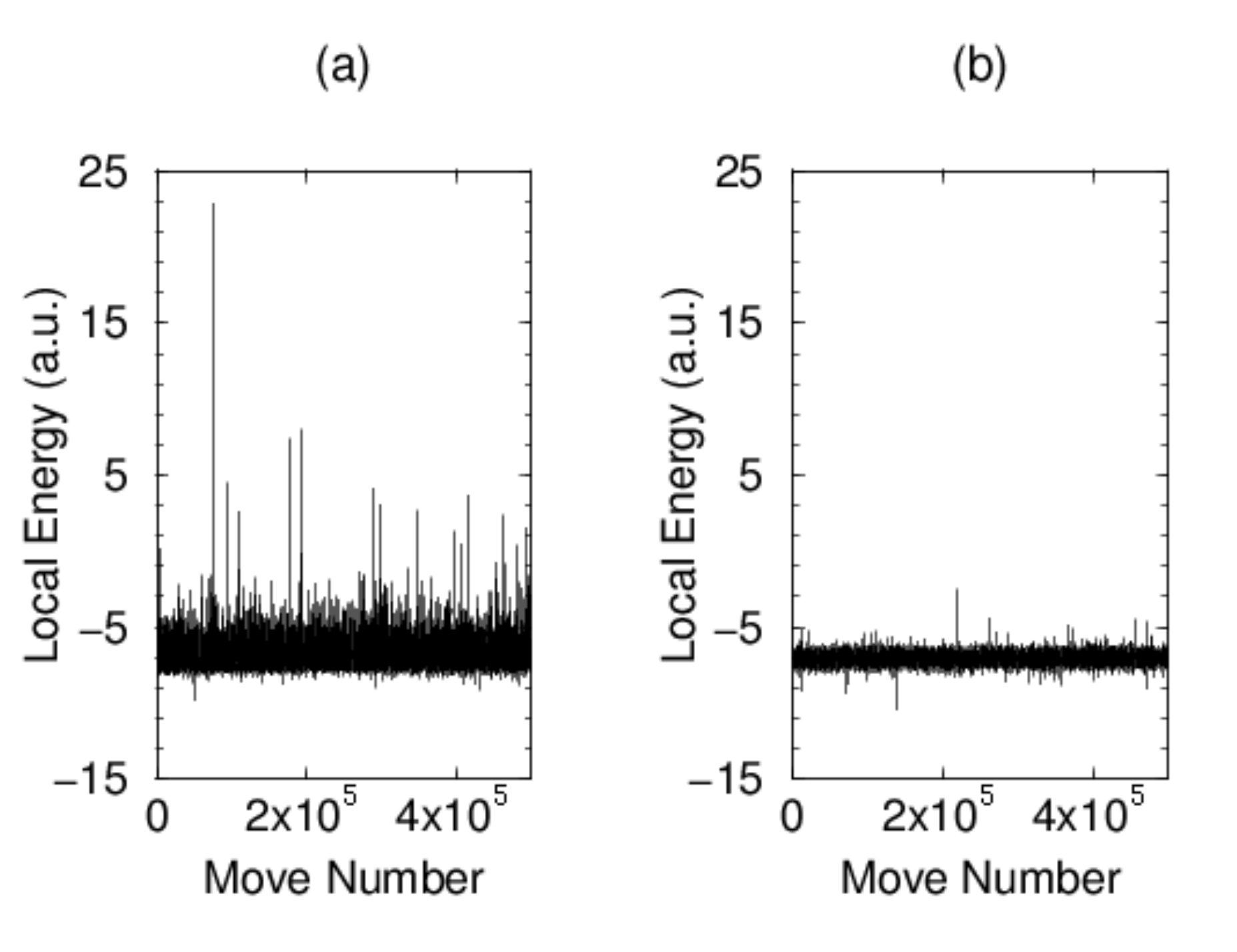}
\caption[]{Local energy of a silane (SiH$_4$) molecule from a VMC calculation
(a) using a Slater-determinant trial wave function and (b) including a Jastrow
factor.}
\label{fig:1}
\end{figure}

Figure \ref{fig:1} shows the local energies generated during two VMC runs for
a silane molecule in which the Si$^{4+}$ and H$^{+}$ ions are described by
smooth pseudopotentials.  In figure \ref{fig:1}(a) the trial wave function
consists of a product of up- and down-spin Slater determinants of molecular
orbitals.  The Kato cusp conditions for electron-electron coalescences are
therefore not satisfied and the local energy shows very large positive spikes
when two electrons are close together.  Figure \ref{fig:1}(b) shows the effect
of adding a Jastrow factor which satisfies the electron-electron cusp
conditions.  The large positive spikes in the local energy are removed and the
mean energy is lowered.  Some small spikes remain, and the frequency and size
of the positive and negative spikes are roughly equal.  These spikes arise
from electrons approaching the nodes of the trial wave function, where the
local kinetic energy diverges positively on one side of the node and
negatively on the other side.

The basic Jastrow factor that we use for systems of electrons and ions
contains the sum of homogeneous, isotropic electron-electron terms $u$,
isotropic electron-nucleus terms $\chi$ centred on the nuclei and isotropic
electron-electron-nucleus terms $f$, also centred on the nuclei
\cite{ndd_newjas}.  We use a Jastrow factor of the form $\exp [J({\bf R})]$,
where
\begin{eqnarray}
\label{eq:basic_J}
J(\{{\bf r}_i\},\{{\bf r}_I\}) = \sum_{i>j}^{N} u(r_{ij}) + \sum_{I=1}^{N_{\rm
ions}} \sum_{i=1}^N \chi_I(r_{iI}) + \sum_{I=1}^{N_{\rm ions}} \sum_{i>j}^{N}
f_I(r_{iI},r_{jI},r_{ij}) \;,
\end{eqnarray}
$N$ is the number of electrons, $N_{\rm ions}$ is the number of ions, ${\bf
r}_{ij} = {\bf r}_{i} - {\bf r}_{j}$, ${\bf r}_{iI} = {\bf r}_{i} - {\bf
r}_{I}$, ${\bf r}_i$ is the position of electron $i$ and ${\bf r}_I$ is the
position of nucleus $I$.  The functions $u$, $\chi$ and $f$ are represented
by power expansions with optimisable coefficients.  Different coefficients are
used for terms involving different spins.  Note that, even if the determinant
part of the Slater-Jastrow wave function is an eigenfunction of the spin
operator $\hat{S}^2$, the use of different coefficients for parallel-spin and
antiparallel-spin pairs of electrons generally leads to a trial wave function
that is not an eigenfunction of $\hat{S}^2$.

When using periodic boundary conditions, we often add a plane-wave term in the
electron-electron separations, $p({\bf r}_{ij})$, which describes similar
sorts of correlation to the $u$ term.  The $u(r_{ij})$ term, however, is cut
off at a distance less than or equal to the Wigner-Seitz radius of the
simulation cell, and the $p$ term adds variational freedom in the corners of
the simulation cell.  Occasionally we add a plane-wave expansion in electron
position, $q({\bf r}_{i})$, and also occasionally add three-body
electron-electron-electron terms.

\begin{figure}[t]
\centering \includegraphics*[width=.7\textwidth]{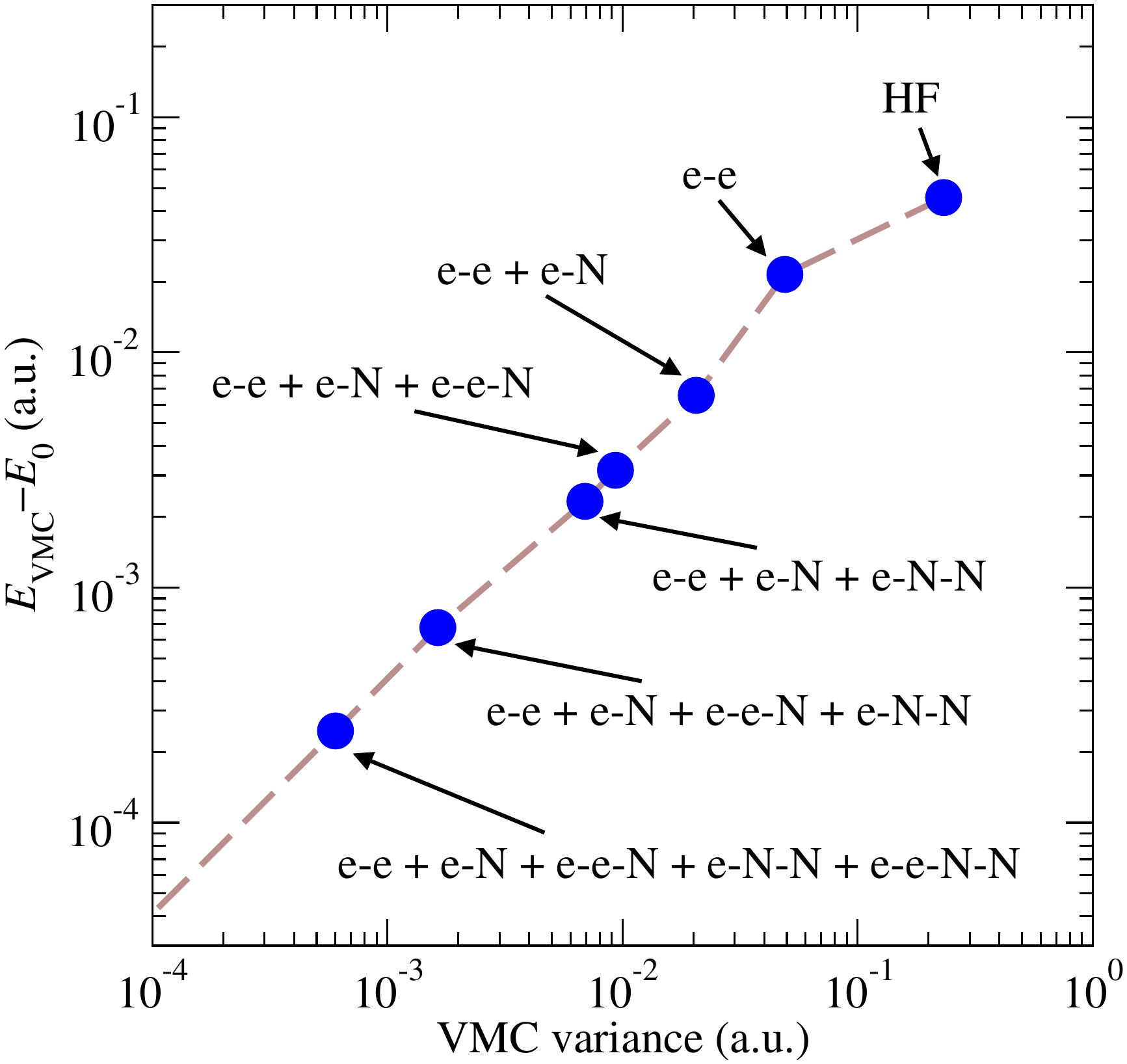}
\caption[]{The difference between the VMC energy and the exact ground state
energy against the variance of the VMC local energies on logarithmic scales
for H${}_2$ at a bond length of 1.397453 a.u.\ obtained using Jastrow factors
of increasing complexity. ``HF'' indicates a wave function consisting of a
molecular orbital obtained from a Hartree-Fock calculation and ``e-e-N''
denotes a term in the Jastrow factor involving the three distances between two
electrons and one proton, \textit{etc}.
}
\label{fig:h2jastrow}
\end{figure}

We have recently developed a more general form of Jastrow factor
\cite{general_jastrow} which allows the inclusion of higher order terms than
those of equation (\ref{eq:basic_J}), such as terms involving the distances
between four or more particles.  An example of the application of such a
Jastrow factor to the H${}_2$ molecule is shown in figure \ref{fig:h2jastrow}.
The molecular orbital was calculated within Hartree-Fock theory and VMC
calculations were performed including Jastrow factors of increasing
complexity.  The Jastrow factor of equation (\ref{eq:basic_J}) includes
electron-nucleus (e-N \textit{etc.}), e-e and e-e-N terms, but the additional
reductions in energy from including the e-N-N and e-e-N-N terms are clearly
visible in figure \ref{fig:h2jastrow}.


\subsection{Pairing wave functions}
\label{subsec:Pairing wave functions}

Slater-Jastrow wave functions are not appropriate for all systems.  For
example, the strongly attractive interaction between electrons and holes
within an effective-mass theory leads to the formation of excitons, which are
not well described by a Slater-Jastrow wave function.  A more appropriate wave
function \cite{depalo_2002} is formed from the antisymmetrised product of
identical electron-hole pairing functions $\psi$, multiplied by a Jastrow
factor,
\begin{equation}
\label{eq:singlet-pairing}
\Psi_{\rm SP}({\bf R}) = e^{J({\bf R})} \det{\left[\psi({\bf
r}_i^{\uparrow},{\bf r}_j^{\downarrow})\right]} \;.
\end{equation}
It is also possible to include additional orbitals for unpaired particles
within this wave function.

\subsection{Multi-determinant wave functions}
\label{subsec:Multi-determinant wave functions}

Multi-determinant expansions have been used with considerable success over
many decades within the quantum chemistry community.  The trial wave function
can be written as
\begin{equation}
\label{eq:multi-slater-jastrow}
\Psi_{\rm MD}({\bf R})\! \! = e^{J({\bf R})} \sum_n c_n \det{\left[
\psi_n({\bf r}_i^{\uparrow})\right]} \det{\left[ \psi_n({\bf
r}_j^{\downarrow})\right]} \;,
\end{equation}
where the $c_n$ are coefficients.  This method provides a systematic approach
to improving the trial wave function, and there have been numerous
applications of multi-determinant trial wave functions in QMC calculations for
small molecules \cite{filippi_1996,schautz_2004,harkless_2006}.  Such trial
wave functions can capture near-degeneracy effects (also known as
\textit{static correlation}).  Multi-determinant wave functions are not in
general suitable for large systems because the number of determinants required
to retrieve a given fraction of the correlation energy increases exponentially
with system size.  An exception to this occurs if only a small region of the
system requires a multi-determinant description.  An example of a DMC
calculation of this type is the study of the electronic states formed by the
strongly interacting dangling bonds at a neutral vacancy in diamond by Hood
\textit{et al.}\ \cite{hood_2003}.

\subsection{Backflow wave functions}
\label{subsec:Backflow wave functions}

\begin{figure}[t]
\centering \includegraphics*[width=.7\textwidth]{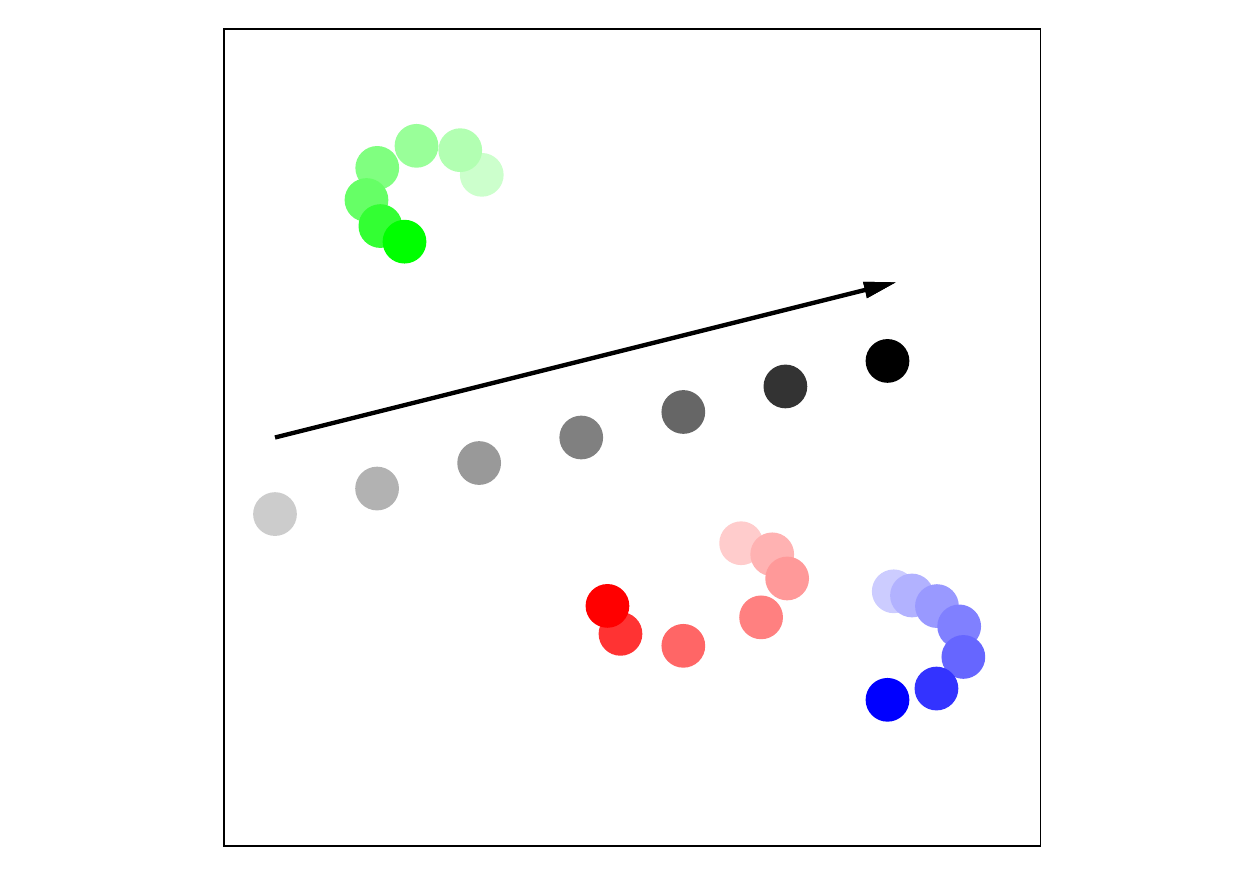}
\caption[]{Effect of the motion of an electron (black, with the arrow showing
the direction of motion) on the backflow-transformed coordinates of three
opposite-spin electrons (red, green and blue).  Circles with the same colour
intensity correspond to the same instant in the motion.
}
\label{fig:bfplot}
\end{figure}

Additional correlation effects can be incorporated in the trial wave function
using backflow transformations \cite{feynman_1954,feynman_1956}.  Consider a
solid ball falling through a classical liquid.  The incompressible liquid is
pushed out of the way and it fills in behind the ball to form a characteristic
flow pattern.  One can imagine that similar correlations occur as a quantum
particle moves through a quantum fluid, as shown in figure \ref{fig:bfplot}.
Much of this correlation can be captured in a Jastrow factor which, however,
preserves the nodal surface of the wave function.  The backflow motion gives
an additional contribution which leaves its imprint on the nodes.  Quantum
backflow was discussed by Feynman and coworkers
\cite{feynman_1954,feynman_1956} for excitations in $^{4}$He and the effective
mass of a $^{3}$He impurity in liquid $^{4}$He.  Backflow wave functions have
been used successfully in QMC studies of liquid He
\cite{schmidt_1981,holzmann_2006}, the electron gas
\cite{kwon_1993,kwon_1998,zong_2002}, hydrogen systems \cite{delaney_2006},
and various inhomogeneous systems
\cite{lopez-rios_2006,drummond_2006b,brown_2007}.

The backflow wave functions we use \cite{lopez-rios_2006} can be written as
\begin{equation}
\label{eq:backflow}
\Psi_{\rm BF}({\bf R}) = e^{J({\bf R})} \det{\left[\psi_i({\bf
r}_i^{\uparrow}+{\bm \xi}_i({\bf R}))\right]} \det{\left[\psi_i({\bf
r}_j^{\downarrow}+{\bm \xi}_j({\bf R}))\right]} \;.
\end{equation}
For a system of $N$ electrons and $N_{\rm ion}$ classical ions we write the
backflow displacement for electron $i$ in the form
\begin{equation}
\label{eq:backflow displacement}
{\bm \xi}_i = \sum_{j\neq i}^N \eta_{ij} {\bf r}_{ij} + \sum_I^{N_{\rm ion}}
\mu_{iI} {\bf r}_{iI} + \sum_{j\neq i}^N \sum_I^{N_{\rm ion}} {\big (}
\Phi_i^{jI} {\bf r}_{ij} + \Theta_i^{jI} {\bf r}_{iI} {\big )} \;.
\end{equation}
In this expression $\eta_{ij}=\eta(r_{ij})$ is a function of electron-electron
separation, $\mu_{iI}=\mu(r_{iI})$ is a function of electron-ion separation,
and $\Phi_i^{jI}=\Phi(r_{iI},r_{jI},r_{ij})$ and
$\Theta_i^{jI}=\Theta(r_{iI},r_{jI},r_{ij})$.  We parameterise the functions
$\eta$, $\mu$, $\Phi$ and $\Theta$ using power expansions with optimisable
coefficients \cite{lopez-rios_2006}.

\subsection{Other wave functions}
\label{subsec:Other wave functions}

The wave function types of equations (\ref{eq:slater-jastrow}),
(\ref{eq:singlet-pairing}), (\ref{eq:multi-slater-jastrow}), and
(\ref{eq:backflow}) can be combined in various ways within the \textsc{casino}
code \cite{casino} so that, for example, it is possible to use
Slater-Jastrow-pairing-backflow wave functions, \textit{etc.}  Of course the
range of possible wave functions could be extended by, for example, including
Pfaffian wave functions \cite{bajdich_2006,bajdich_2008}, \textit{etc.}

\section{Optimisation of trial wave functions}
\label{sec:optimise_psi_trial}

Optimising trial wave functions is a very important part of QMC calculations
which can consume large amounts of human and computing resources.  With modern
stochastic methods it is possible to optimise hundreds or even thousands of
parameters in the wave function.  The parameters which can be optimised
include those in the Jastrow factor, the coefficients of determinants in a
multi-determinant wave function, the parameters in the backflow functions and
the parameters in single-particle and pairing orbitals.

The trial wave function used in a DMC calculation should ideally be optimised
within DMC, but reliable and efficient methods to achieve this are still under
development \cite{luchow_2007,reboredo_2009}.  Minimisation of the DMC energy
has been performed ``by hand'' for small numbers of parameters
\cite{drummond_2004,drummond_2008_2d}.  Wave function optimisation within
\textsc{casino} is performed by minimising the VMC energy or its variance.

Optimising wave functions by minimising the variance of the energy is an old
idea dating back to the 1930s.  The first application within Monte Carlo
methods may have been by Conroy \cite{conroy_1964}, but the method was
popularised within QMC by the work of Umrigar and coworkers
\cite{umrigar_1988}.  It is now generally believed that it is better to
minimise the VMC energy than its variance, but it has proved more difficult to
develop robust and efficient algorithms for this purpose.  Since the trial
wave function forms used cannot generally represent energy eigenstates
exactly, except in trivial cases, the minima in the energy and variance do not
coincide.  Energy minimisation should therefore produce lower VMC energies,
and although it does not necessarily follow that it produces lower DMC
energies, experience indicates that, more often than not, it does.

\subsection{Variance minimisation}
\label{subsec:variance minimisation}

The variance of the VMC energy is
\begin{equation}
\label{eq:sigma^2_vmc}
\sigma^2({\bm \alpha}) =  \frac{\int [\Psi_{\rm T}^{{\bm \alpha}}({\bf R})]^2
[E_{\rm L}^{{\bm \alpha}}({\bf R}) -  E_{\rm V}^{{\bm \alpha}}]^2 \, d{\bf
R}}{\int [\Psi_{\rm T}^{{\bm \alpha}}({\bf R})]^2 \, d{\bf R}} \;,
\end{equation}
where ${{\bm \alpha}}$ denotes the set of variable parameters.  The minimum
possible value of $\sigma^2({{\bm \alpha}})$ is zero, which is obtained if and
only if $\Psi_{\rm T}^{{\bm \alpha}}$ is an exact eigenstate of $\hat{H}$.  In
practice the trial wave function forms used are incapable of representing the
exact eigenstates.  Nevertheless, the minimum value of $\sigma^2({{\bm
\alpha}})$ is still expected to correspond to a reasonable set of wave
function parameters.

Minimisation of $\sigma^2({{\bm \alpha}})$ is carried out via a correlated
sampling approach in which a set of configurations distributed according to
$[\Psi_{\rm T}^{{\bm \alpha}_0}]^2$ is generated, where ${{\bm \alpha}_0}$ is
an initial set of parameter values \cite{dewing_2002}.  $\sigma^2({{\bm
\alpha}})$ is then evaluated as
\begin{equation}
\label{eq:sigma^2_vmc correlated sampling}
\sigma^2({{\bm \alpha}}) = \frac{\int [\Psi_{\rm T}^{{\bm \alpha}_0}]^2 \;
w_{{\bm \alpha}_0}^{{\bm \alpha}} \; [E_{\rm L}^{{\bm \alpha}} - E_{\rm
V}^{{\bm \alpha}}]^2 \, d{\bf R}}{\int [\Psi_{\rm T}^{{\bm \alpha}_0}]^2 \;
w_{{\bm \alpha}_0}^{{\bm \alpha}} \, d{\bf R}} \;,
\end{equation}
where the integrals contain weights, $w_{{\bm \alpha}_0}^{{\bm \alpha}}$,
    given by
\begin{equation}
\label{eq:W}
w_{{\bm \alpha}_0}^{{\bm \alpha}}({\bf R}) = \frac{[\Psi_{\rm T}^{{\bm
  \alpha}}]^2} {[\Psi_{\rm T}^{{\bm \alpha}_0}]^2} \;,
\end{equation}
and $E_{\rm V}$ is evaluated using
\begin{equation}
\label{eq:vmc energy correlated sampling}
E_{\rm V} =  \frac{ \int  [\Psi_{\rm T}^{{\bm \alpha}_0}]^2 \; w_{{\bm
\alpha}_0}^{{\bm \alpha}} \;  E_{\rm L}^{{\bm \alpha}} \, d{\bf R} } {\int
[\Psi_{\rm T}^{{\bm \alpha}_0}]^2 \;  w_{{\bm \alpha}_0}^{{\bm \alpha}} \,
d{\bf R}} \;.
\end{equation}

After generating the initial set of configurations, the optimisation proceeds
using standard techniques to locate the new parameter values which minimise
$\sigma^2({{\bm \alpha}})$.  With perfect sampling $\sigma^2({{\bm \alpha}})$
is independent of the initial parameter values ${{\bm \alpha}_0}$.  For real
(finite) sampling, however, one runs into problems because the values of
$w_{{\bm \alpha}_0}^{{\bm \alpha}}$ for different configurations can vary by
many orders of magnitude if ${{\bm \alpha}}$ and ${{\bm \alpha}_0}$ differ
substantially.  During the minimisation procedure a few configurations (often
only one) acquire very large weights and the estimate of the variance is
reduced almost to zero by a poor set of parameter values.  This optimisation
scheme is therefore often unstable, and in practice modified versions of it
are used.


The above scheme can be made much more stable by altering the weights $w_{{\bm
\alpha}_0}^{{\bm \alpha}}$.  A robust procedure is to set all the weights
$w_{{\bm \alpha}_0}^{{\bm \alpha}}$ in equation (\ref{eq:sigma^2_vmc
correlated sampling}) to unity, which is reasonable because the minimum value
of $\sigma^2({{\bm \alpha}}) = 0$ is still obtained only if $E_{\rm L}({\bf
R})$ is a constant independent of ${\bf R}$, which holds only for eigenstates
of the Hamiltonian.  We call this the ``unreweighted variance'' minimisation
method.  The procedure is cycled until the parameters converge to their
optimal values (within the statistical noise).  For a number of model systems
it was found that the trial wave functions generated by unreweighted variance
minimisation iterated to self-consistency have a lower variational energy than
wave functions optimised by reweighted variance minimisation
\cite{drummond_2005_min}.

If the Jastrow factor of equation (\ref{eq:basic_J}) can be written in the form
\begin{equation}
\label{eq:linear Jastrow factor}
J({\bf R}) = \sum_n \alpha_n f_n({\bf R}) \;,
\end{equation}
then it is possible to simplify the calculation of the variance of the VMC
energy \cite{moroni_1995b,drummond_2005_min}.  It can be shown that the
unreweighted variance is a quartic function of the linear parameters
$\alpha_n$ \cite{drummond_2005_min}.  This has two advantages: ($i$) the
unreweighted variance can be evaluated extremely rapidly at a cost which
depends only on the number of parameters and is independent of the number of
particles; and ($ii$) the unreweighted variance along a line in parameter
space is a quartic polynomial.  This is useful because it allows the exact
global minimum of the unreweighted variance along the line to be computed
analytically by solving the cubic equation obtained by setting the derivative
equal to zero.

The unreweighted variance minimisation method works well for optimising
Jastrow factors, but it often performs poorly when parameters which alter the
nodal surface of $\Psi_{\rm T}$ are optimised.  The problem is that the local
energy $E_{\rm L}$ generally diverges for a configuration on the nodal
surface.  As the parameter values are changed during a minimisation cycle the
nodal surface can move through a configuration, resulting in a very large
(positive or negative) value of $E_{\rm L}$, which adversely affects the
optimisation.  Such an effect would not occur when using the weights $w_{{\bm
\alpha}_0}^{{\bm \alpha}}$ because they go to zero on the nodal surface.  We
have developed two schemes which solve this problem. In the first scheme we
limit the weights by replacing them with ${\rm min}(w_{{\bm \alpha}_0}^{{\bm
\alpha}},W)$, so that the weight goes to zero on the nodal surface but can
never become larger than a chosen value $W$.  In the second scheme we use a
weight which goes smoothly to zero as $E_{\rm L}$ deviates from an estimate of
the energy.

Unreweighted variance minimisation belongs to a wider class of wave-function
optimisation methods which are based on minimising a measure of the spread of
the set of local energies.  Another measure of spread that we have used with
considerable success for wave-function optimisation is the mean absolute
deviation of the local energies of a set of configurations from the median
energy,
\begin{equation}
\label{eq:madmin}
{\cal M} = \frac{\int [\Psi_{\rm T}^{{\bm \alpha}_0}({\bf R})]^2 |E_{\rm
    L}^{{\bm \alpha}}({\bf R}) - E_{\rm m}^{{\bm \alpha}}| \, d{\bf R}}{\int
  [\Psi_{\rm T}^{{\bm \alpha}_0}({\bf R})]^2 \, d{\bf R}} \;.
\end{equation}
In this expression, $E_{\rm m}^{{\bm \alpha}}$ is the median value of the
local energies evaluated with the parameter values ${{\bm \alpha}}$.  This is
useful for optimising parameters that affect the nodal surface, because
outlying local energies are less significant.

\subsection{Energy minimisation}
\label{subsec:energy minimisation}

A well-known method for finding approximations to the eigenstates of a
Hamiltonian is to express the wave function as a linear combination of basis
states $g_i$,
\begin{equation}
\label{eq:linear parameters}
\Psi_{\rm T}({\bf R}) = \sum_{i=1}^p \beta_i \, g_i({\bf R}) \;,
\end{equation}
calculate the matrix elements $H_{ij} = \langle g_i| \hat{H}| g_j \rangle$ and
$S_{ij} = \langle g_i| g_j \rangle$, and solve the two-sided eigenproblem
$\sum_j H_{ij}\beta_j = E \sum_j S_{ij}\beta_j$ by standard diagonalisation
techniques.  One can also do this in QMC \cite{riley_2003}, although the
statistical noise in the matrix elements leads to slow convergence with
respect to the number of configurations used to evaluate the integrals.

Nightingale and Melik-Alaverdian \cite{nightingale_emin} reformulated the
diagonalisation procedure as a least-squares fit rather than integral
evaluation, which leads to much faster convergence with the number of
configurations.  Let us assume that the set $\{g_i\}$ spans an invariant
subspace of $\hat{H}$, which means that the result of acting $\hat{H}$ on any
member of the set $\{g_i\}$ can be expressed as a linear combination of the
$\{g_i\}$, \textit{i.e.},
\begin{equation}
\label{eq:invariant subspace}
\hat{H} g_i({\bf R}) = \sum_{i=1}^p {\cal{E}}_{ij} g_j({\bf R}) \;\;\;\;\;
\forall \; i \;.
\end{equation}
The eigenstates and associated eigenvalues of $\hat{H}$ can then be obtained
by diagonalising the matrix ${\cal{E}}_{ij}$.  Within a Monte Carlo approach
we could evaluate the $g_i({\bf R})$ and $\hat{H} g_i({\bf R})$ for $p$
uncorrelated configurations generated by a VMC calculation and solve the
resulting set of linear equations for the ${\cal{E}}_{ij}$.  For problems of
interest, however, the assumption that the set $\{g_i\}$ span an invariant
subspace of $\hat{H}$ does not hold and there exists no set of
${\cal{E}}_{ij}$ which solves equation (\ref{eq:invariant subspace}).  If we
took $p$ configurations and solved the set of $p$ linear equations, the values
of ${\cal{E}}_{ij}$ would depend on which configurations had been chosen.  To
overcome this problem, a number of configurations $M \gg p$ is sampled to
obtain an overdetermined set of equations which can be solved in a
least-squares sense using singular value decomposition.  In fact Nightingale
and Melik-Alaverdian recommended that equation (\ref{eq:invariant subspace})
be divided by $\Psi_{\rm T}({\bf R})$ so that in the limit of perfect sampling
the scheme corresponds precisely to standard diagonalisation.

The method of Nightingale and Melik-Alaverdian works very well for linear
variational parameters as in equation (\ref{eq:linear parameters}).  The
natural generalisation to parameters which appear non-linearly in $\Psi_{\rm
T}$ is to consider the basis of the initial trial wave function ($g_0 =
\Psi_{\rm T}$) and its derivatives with respect to the variable parameters,
\begin{equation}
\label{eq:non-linear parameters}
g_i = \left. \frac{\partial \Psi_{\rm T}}{\partial \beta_i}
\right|_{\beta_i^0} \;.
\end{equation}
In its simplest form this algorithm turns out to be highly unstable because
the first-order approximation in equation (\ref{eq:non-linear parameters}) is
often inadequate.  Umrigar and coworkers \cite{umrigar_emin,toulouse_emin}
showed how this method can be stabilised.  The details of the stabilisation
procedures are quite involved and we refer the reader to the original papers
\cite{umrigar_emin,toulouse_emin} for the details.  The stabilised algorithm
works well and is quite robust.  The VMC energies given by this method are
usually lower than those obtained from any of the variance-based algorithms
described in section \ref{subsec:variance minimisation}, although the
difference is often small.

\section{QMC calculations within periodic boundary conditions}
\label{sec:pbc}

QMC calculations for extended systems may be performed using cluster models or
periodic boundary conditions, just as in other techniques.  Periodic boundary
conditions are preferred because they give smaller finite size effects.  One
can also use the standard supercell approach for systems which lack
three-dimensional periodicity where a cell containing, for example, a point
defect and a small part of the host crystal, are repeated periodically
throughout space.  Just as in other electronic structure methods, one must
ensure that the supercell is large enough for the interactions between defects
in different supercells to be small.

When using standard single-particle-like theories within periodic boundary
conditions such as density functional theory, the charge density and
potentials are taken to have the periodicity of a chosen unit cell or
supercell.  The single particle orbitals can then be chosen to obey Bloch's
theorem and the results for the infinite system are obtained by summing
quantities obtained from the different Bloch wave vectors within the first
Brillouin zone.  This procedure can also be applied within HF calculations,
although the Coulomb interaction couples the Bloch wave vectors in pairs.
The situation with the many-particle wave functions described in
section \ref{sec:psi_trial} is somewhat different.  Although the
many-particle wave function satisfies Bloch theorems
\cite{kpoints_1,kpoints_2}, it is not possible to perform a
many-particle calculation using a set of ${\bf k}$-points; one has to
perform it at a single ${\bf k}$-point.  A single ${\bf k}$-point
normally gives a poor representation of the infinite-system result, so
that one either chooses a larger non-primitive simulation cell, or
averages over the results of QMC calculations at a set of different
${\bf k}$-points \cite{lin_2001}, or both.


Many-body techniques such as QMC also suffer from finite size errors arising
from long-ranged interactions, most notably the Coulomb interaction.  Coulomb
interactions are normally included within periodic boundary conditions
calculations using the Ewald interaction.  Long-ranged interactions induce
long-ranged exchange-correlation interactions, and if the simulation cell is
not large enough these effects are described incorrectly.  Such effects are
absent in local DFT calculations because the interaction energy is written in
terms of the electronic charge density, but HF calculations show very strong
effects of this kind and various ways to accelerate the convergence have been
developed.  The finite size effects arising from the long-ranged interaction
can be divided into potential and kinetic energy contributions
\cite{fin_chiesa,ndd_fin}.  The potential energy component can be removed from
the calculations by replacing the Ewald interaction by the so-called model
periodic Coulomb (MPC) interaction \cite{fraser_1996,finsize,finlong}.  Recent
work has added substantially to our understanding of finite size effects, and
theoretical expressions have been derived for them \cite{fin_chiesa,ndd_fin},
but at the moment it seems that they cannot entirely replace extrapolation
procedures.

Kwee \textit{et al.\@} \cite{kwee_2008} have developed an alternative approach
for estimating finite size errors in QMC calculations.  DMC results for the
three-dimensional HEG are used to obtain a system-size-dependent local density
approximation (LDA) functional.  The correction to the total energy is given
by the difference between the DFT energies for the finite-sized and infinite
systems.  This approach appears promising, although it does rely on the LDA
giving a reasonable description of the system.

\section{Pseudopotentials in QMC calculations}
\label{sec:pseudopots}

The computational cost of a DMC calculation increases with the atomic number
$Z$ of the atoms as roughly $Z^{5.5}$ \cite{ceperley_1986,ma_2005} which makes
calculations with $Z>10$ extremely expensive.  This problem can be solved by
using pseudopotentials to represent the effect of the atomic core on the
valence electrons.  The use of non-local pseudopotentials within VMC is quite
straightforward \cite{fahy_prl,fahy_prb}, but DMC poses an additional problem
because the use of a non-local potential is incompatible with the fixed-node
boundary condition.  To circumvent this difficulty an additional approximation
is made.  In the ``locality approximation'' \cite{mitas_1991} the non-local
part of the pseudopotential $\hat{V}_{\rm nl}$ is taken to act on the trial
wave function rather than the DMC wave function, \textit{i.e.}, $\hat{V}_{\rm
nl}$ is replaced by $\Psi_{\rm T}^{-1} \hat{V}_{\rm nl} \Psi_{\rm T}$.  The
leading-order error term in the locality approximation is proportional to
$(\Psi_{\rm T} - \phi_0)^2$ \cite{mitas_1991}, where $\phi_0$ is the exact
fixed-node ground state wave function, although it can be of either sign, so
that the variational property of the algorithm is lost.  Casula \textit{et
al.\@} \cite{casula_2005,casula_2006} have introduced a fully variational
``semi-localisation'' scheme for dealing with non-local pseudopotentials
within DMC, which also shows superior numerical stability to the locality
approximation.

Currently it is not possible to generate pseudopotentials entirely within a
QMC framework, and therefore they are obtained from other sources.  There is
evidence that HF theory provides better pseudopotentials than DFT for use
within QMC calculations \cite{greeff_1998}, and we have developed smooth
relativistic HF pseudopotentials for H to Ba and Lu to Hg, which are suitable
for use in QMC calculations \cite{trail_2005_1,trail_2005_2,casino_page}.
Another set of pseudopotentials for use in QMC calculations has been developed
by Burkatzki \textit{et al.\@} \cite{burkatzki07}.  In the few cases where
reliable tests have been performed \cite{trail_2008,santra_2008}, the
pseudopotentials of references \cite{trail_2005_1,trail_2005_2,casino_page} and
those of \cite{burkatzki07} have produced almost identical results, although
those of references \cite{trail_2005_1,trail_2005_2,casino_page} are a little
more efficient as they have smaller core radii.

\section{DMC calculations for excited states}
\label{sec:excited_states}

The fixed-node DMC algorithm is useful for studying excited states
because it gives the exact excited-state energy if the nodal surface
of the trial wave function matches that of the exact excited state
and it gives an approximation to the excited-state energy if a trial
wave function with an approximate nodal surface is used.

This can be proved as follows.  The local energy calculated with the
exact excited-state wave function is equal to the exact excited-state
energy throughout configuration space, and, by definition, the wave
function is zero at the nodal surface and nowhere else.  Hence within
each nodal pocket the exact excited-state wave function is the
ground-state solution of the Schr\"odinger equation subject to the
boundary condition of being zero on the pocket boundary.  Therefore
the ground-state pocket eigenvalues are all equal to the exact
excited-state energy, and the fixed-node DMC algorithm indeed gives
the exact excited-state energy.

An important difference from the ground state case is that the
existence of a variational principle for excited state energies cannot
in general be guaranteed, and it depends on the symmetry of the trial
wave function \cite{foulkes_1999}. In practice DMC works quite well
for excited states
\cite{williamson_1998,towler_2000,porter_2001a,porter_2001b,williamson_2002,drummond_2005_dia,bande_2006}.
Ceperley and Bernu \cite{ceperley_1988} have devised a method which
combines DMC and the variational principle to calculate the
eigenvalues of several different excited states simultaneously.
However, this method suffers from stability problems in large systems.

\section{Scaling of computational effort with system size}
\label{sec:scaling}

Over the accessible range of system sizes, the computational cost of a
single configuration move in a VMC or DMC calculation is usually
determined by the time taken to evaluate each of the ${\cal O}(N)$
orbitals in the Slater part of the wave function at each of the $N$
electron positions \cite{foulkes_2001}.  If the delocalised orbitals
are expanded in localised basis functions then the time taken to move
a configuration scales as ${\cal O}(N^2)$.  However, the number of
configuration moves required to achieve a given error bar on the total
energy grows as ${\cal O}(N)$, because the variance of the energy is
proportional to the system size.  Hence the time taken to evaluate the
total energy to within a given statistical error bar scales as ${\cal
O}(N^3)$.  (Note that the time taken to evaluate the Slater
determinants during the run scales as ${\cal O}(N^4)$, but with a
small prefactor.  In fact, for the DMC method the scaling with system
size is ultimately exponential due to correlations within the
configuration population \cite{note_exp_scaling}.)

The scaling of the QMC methods can be improved by using localised
orbitals, so that the number of nonzero orbitals to be evaluated at
each electron position is independent of the system size
\cite{williamson_2001,alfe_2004}.  In this case the CPU time required
to achieve a given error bar on the total energy scales as ${\cal
O}(N^2)$ over the relevant range of system sizes.  To maximise the
localisation of the orbitals, the orthogonality constraint can be
dropped, for it is irrelevant in QMC\@.  However, it is not possible
to ``cheat'' on the size of the orbital localisation regions in QMC,
because this would compromise the high accuracy of the method.  (The
use of localised orbitals enables the use of sparse linear algebra to
compute the Slater determinants, improving the scaling of this part of
the algorithm by a factor of $N$ as well.)

In calculations of the energy per particle of a periodic crystal the
number of moves required to achieve a given error per particle falls
off as ${\cal O}(N^{-1})$.  Hence the CPU time required to achieve a
given error bar on the energy per particle increases as ${\cal O}(N)$
in the standard algorithm and is roughly independent of the system
size when localised orbitals are used.

\section{Sources of error and statistical analysis}
\label{subsec:errors}

\subsection{Sources of error in DMC calculations}

\begin{figure}[ht]
\centering
\includegraphics*[width=.7\textwidth]{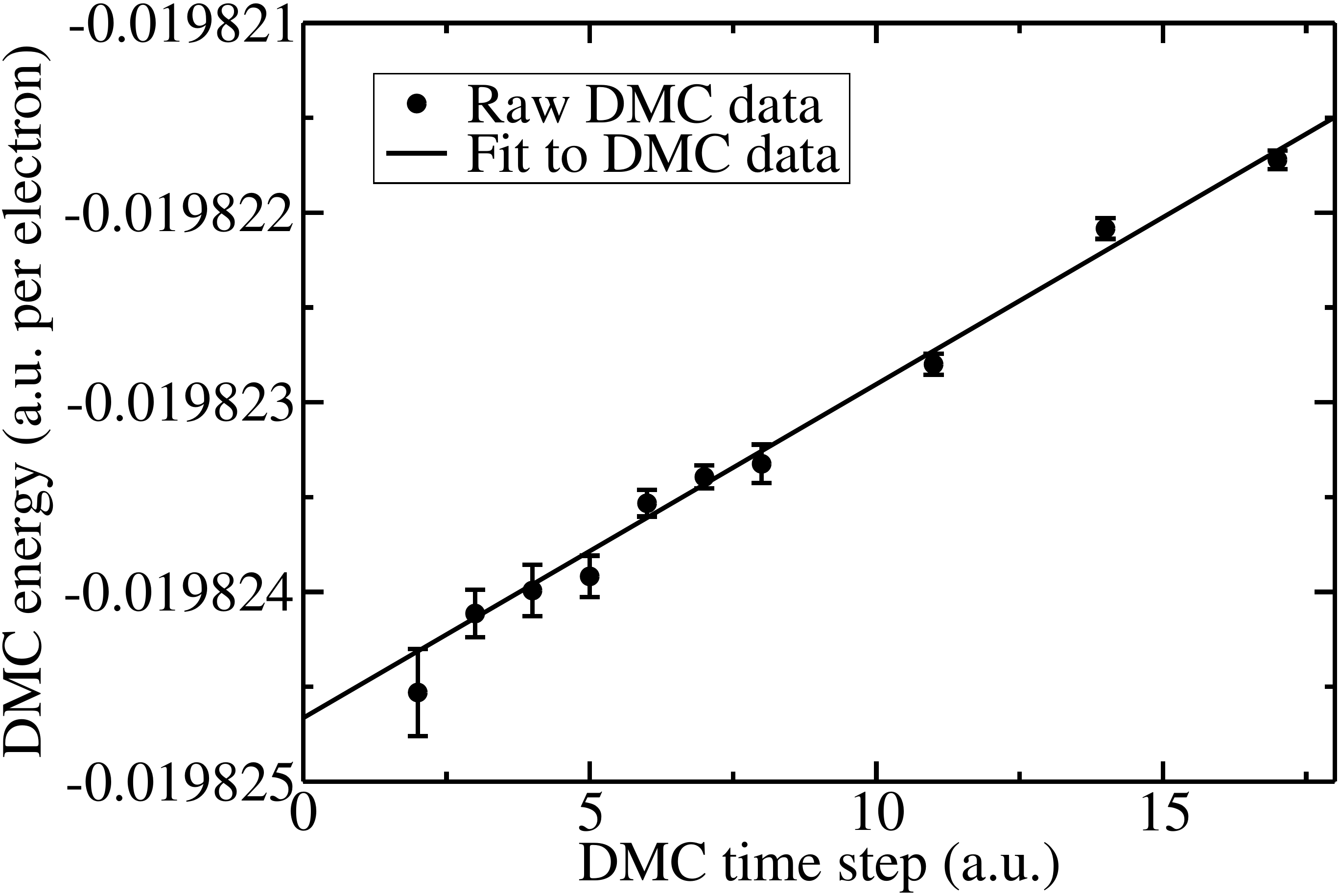}
\caption[]{DMC energy against time step for a 64-electron ferromagnetic 2D
hexagonal Wigner crystal at density parameter $r_s=50$ a.u.\ with a
Slater-Jastrow wave function.  The solid line is a linear fit to the data.
}
\label{fig:time_step_errors}
\end{figure}

The potential sources of errors in DMC calculations may be summarised as
follows.
\begin{itemize}
\item Statistical errors.  The standard error in the mean is proportional to
$1/\sqrt{M}$, where $M$ is the number of particles moves.  It therefore costs
a factor of 100 in computer time to reduce the statistical error bars by a
factor of 10.  On the other hand, a random error is much better than a
systematic one as its size can normally be reliably estimated.
\item Fixed-node error.  This is the central approximation of the DMC
technique, and is normally the limiting factor in the accuracy of the results.
\item Time-step bias.  The short time approximation leads to a bias in the $f$
distribution and hence in expectation values.  This bias is often significant
and can be of either sign, but it can be largely removed by performing
calculations for different time steps and extrapolating to zero time step or
by simply choosing a small enough time step.  An example of time-step
extrapolation is shown in figure \ref{fig:time_step_errors}.
\item Population control bias.  The $f$ distribution is represented by a
finite population of configurations which fluctuates due to branching.  The
population may be controlled in various ways, but this introduces a population
control bias which is positive and falls off as the reciprocal of the
population.  In practice the population control bias is normally so small that
it is difficult to detect \cite{umrigar_1993,drummond_2004}.
\item Finite size errors within periodic boundary conditions calculations.  It
is important to correct for finite size effects carefully, as mentioned in
section \ref{sec:pbc}.
\item The pseudopotential approximation inevitably introduces errors.  In DMC
there is an additional error arising from the localisation \cite{mitas_1991}
or semi-localisation \cite{casula_2006} of the non-local pseudopotential
operator.  The localisation error appears to be quite small in the cases for
which it has been tested \cite{drummond_2006b}.
\end{itemize}

\subsection{Practical methods for handling statistical errors in QMC results}
\label{subsec:statistical errors}

\begin{figure}[ht]
\centering \includegraphics*[width=.7\textwidth]{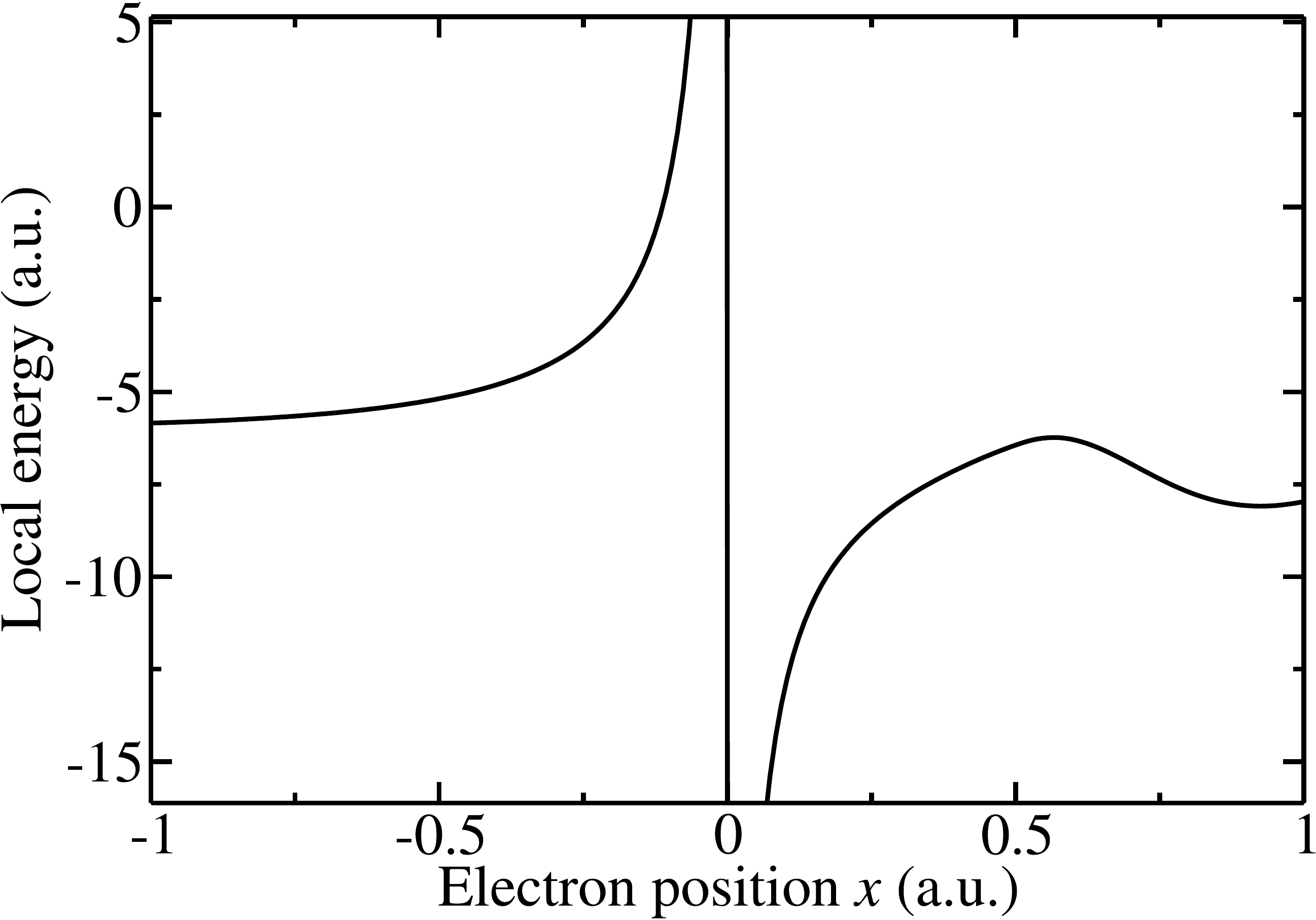}
\caption[]{Variation in the local energy $E_{\rm L}$ of a silane (SiH$_4$)
molecule as an electron moves through the nodal surface at $x=0$.  The local
energy diverges as $1/x$.}
\label{fig:local_energy}
\end{figure}

Two main practical problems are encountered when dealing with errors in the
QMC data: the data are serially correlated and the underlying probability
distributions are non-Gaussian.  The probability distribution of the local
energies has $|E-E_0|^{-4}$ tails, where $E_0$ is a constant.  These tails
arise from singularities in the local energy such as the divergence at the
nodal surface \cite{trail_2005_1,trail_2005_2}, as shown in figure
\ref{fig:local_energy}.  In consequence, although the mean energy and its
variance are well defined, the variance of the variance is infinity.  For
other quantities the problem may be even more severe; for example, the
probability distributions for the Pulay terms in the forces described in
section \ref{subsec:forces} decay as $|F-F_0|^{-5/2}$, so that the variance of
the force is infinity \cite{badinski_2009}.  Reasonably robust estimates of
the errors can still be made, although it has to be accepted that they are not
as well founded as for Gaussian statistics.

\begin{figure}[ht]
\centering \includegraphics*[width=.7\textwidth]{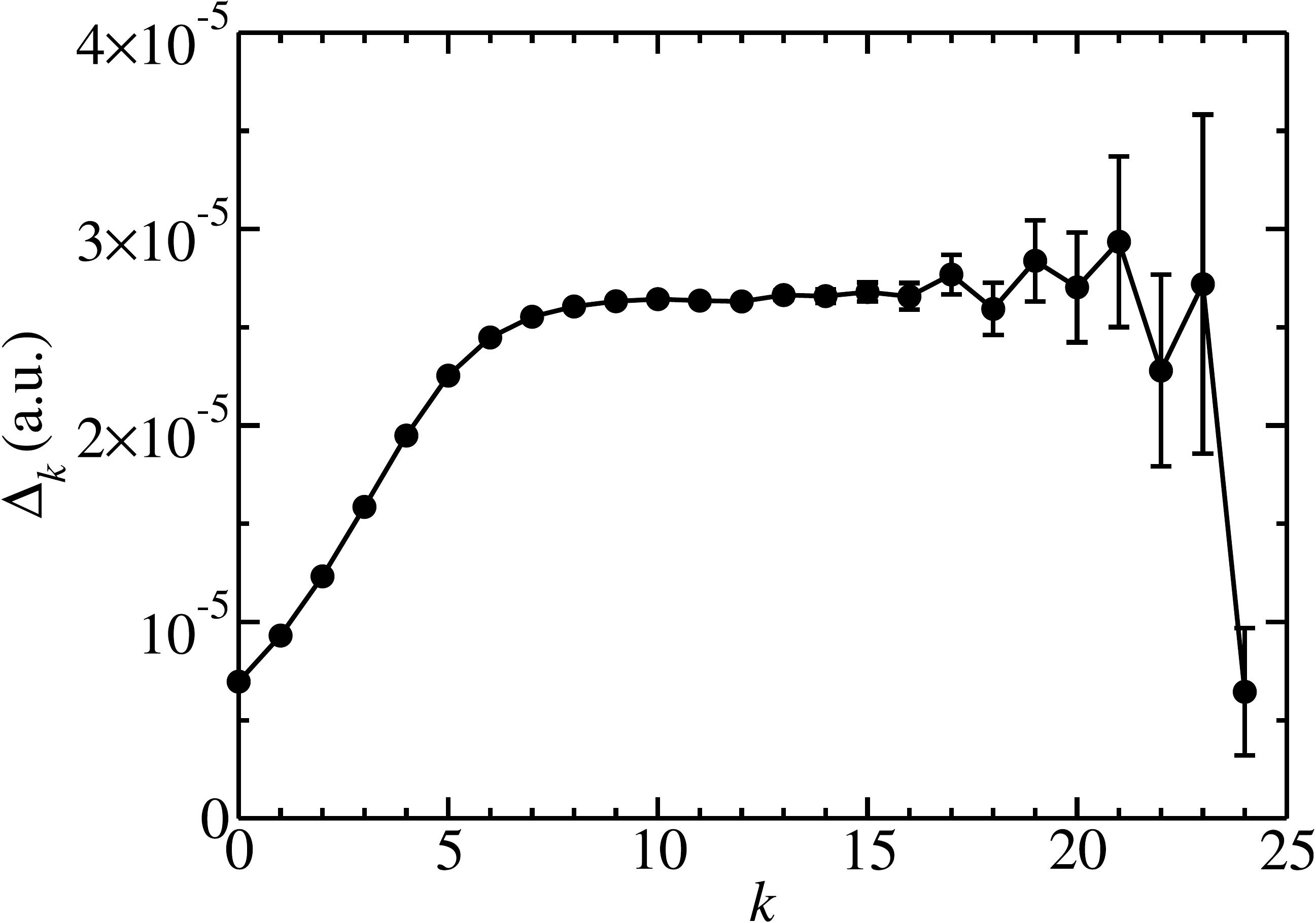}
\caption[]{Blocking analysis of data for an (all-electron) lithium atom.  The
blocking analysis indicates that the true standard error in the mean is about
$\Delta = 2.6 \times 10^{-5}$ a.u., which is reached at about blocking
transformation $k=10$, while the raw value is $\Delta_0 = 7.0 \times 10^{-6}$
a.u.}
\label{fig:reblock}
\end{figure}

The data produced by VMC and DMC calculations are correlated from one step to
the next.  The problem is very important in DMC because short time steps are
used to reduce the effect of the approximation in the Green's function.  The
simulation effectively produces only one independent data point per
correlation time, so that the estimate of the statistical error obtained on
the assumption that the data points are independent is too small.  We use the
``blocking method'' to obtain an estimate of the error.  In this approach
adjacent data points are averaged to form block averages
\cite{Flyvbjerg_1989}.  This procedure is carried out recursively so that
after each blocking transformation the number of data points is reduced by one
half.  An example of blocking is shown in figure \ref{fig:reblock}.  The
computed value of the standard error $\Delta_k$ increases with the number of
blocking transformations $k$ until a limiting value is reached when the block
length starts to exceed the correlation time.  The standard error in the mean
is estimated by the value of $\Delta$ on the plateau.  Because the sizes of
the error bars on QMC expectation values are themselves approximate estimates,
apparent outliers in QMC data can be more common than one might expect on the
basis of Gaussian statistics.

\section{Evaluating other expectation values}
\label{sec:other expectation values}

As mentioned in section \ref{sec:introduction}, VMC and DMC can be used to
calculate expectation values of many time-independent operators, not just the
Hamiltonian.  Typical quantities of interest are particle densities, pair
correlation functions and one- and two-body density matrices, all of which
can be evaluated using the \textsc{casino} code.  It is not possible to obtain
unbiased expectation values directly from the DMC distribution, $f({\bf R})$,
for operators which do not commute with the Hamiltonian (which includes all of
the quantities mentioned in the previous sentence).  Unbiased (within the
fixed-node approximation) estimates can be obtained as pure expectation values,
\begin{eqnarray}
\label{eq:pure expectation value}
\langle \hat{A} \rangle & = & \frac{\int \phi_0({\bf R}) \hat{A} \phi_0({\bf
R}) \, d{\bf R}}{\int \phi_0^2({\bf R}) \, d{\bf R}} \;.
\end{eqnarray}
Pure expectation values can be obtained using a variety of methods: the
approximate (but often very accurate) extrapolation technique
\cite{ceperley_1986b}, the future walking technique
\cite{liu_1974,barnett_1991}, which is formally exact but statistically poorly
behaved, and the reptation QMC technique of Baroni and Moroni
\cite{baroni_1999}, which is formally exact and well behaved, but quite
expensive.  The extrapolation technique can be used for any operator, but the
future walking and reptation techniques are limited to spatially local
multiplicative operators.

Here we shall illustrate the use of the extrapolation technique
\cite{ceperley_1986b} to calculate the charge density of a Wigner crystal. The
pure estimate of the charge density $\rho$ is approximated as
\begin{eqnarray}
\label{eq:extrapolation}
\rho_{\rm ext} \simeq 2\rho_{\rm DMC}-\rho_{\rm VMC}.
\end{eqnarray}
The errors in both the VMC and DMC charge densities $\rho_{\rm VMC}$ and
$\rho_{\rm DMC}$ are linear in the error in the trial wave function, but the
error in the extrapolated estimate $\rho_{\rm ext}$ is quadratic in the error
in the wave function.

\begin{figure}[ht]
\centering \includegraphics*[width=.7\textwidth]{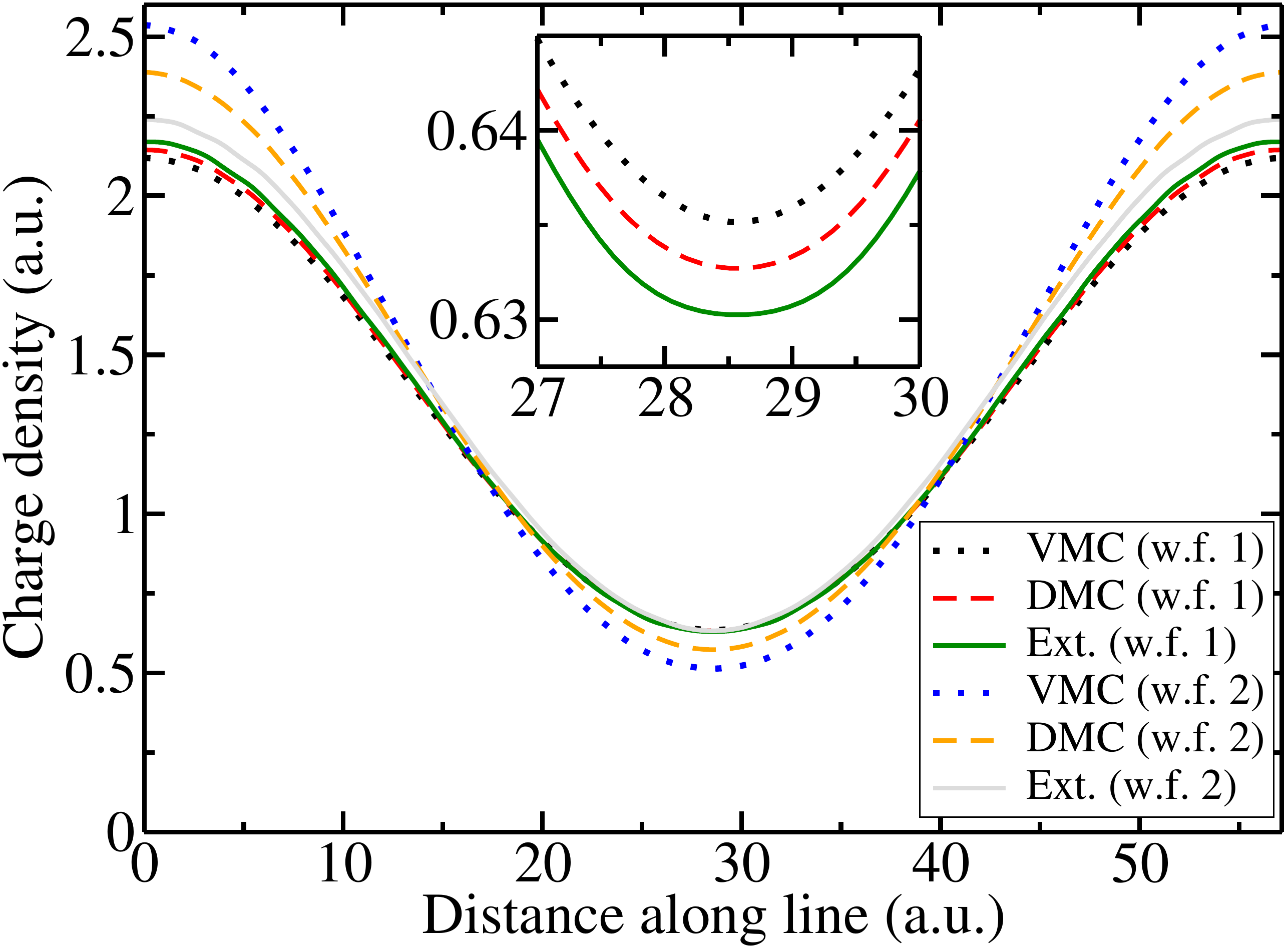}
\caption[]{Charge density of a triangular antiferromagnetic Wigner crystal at
density parameter $r_s=30$ a.u., plotted along a line between a pair of
nearest-neighbour lattice sites.  Two different wave functions are used: wave
function 1 was optimised by variance minimisation, while wave function 2 was
optimised by energy minimisation.  The inset shows the extrapolation with wave
function 1 at the minimum in greater detail.}
\label{fig:crys_cden}
\end{figure}

At low densities the HEG freezes into a Wigner crystal to minimise the
electrostatic repulsion between electrons.  The charge density of a 2D Wigner
crystal \cite{drummond_2008_2d,drummond_2009_2d} close to the crystallisation
density is shown in figure \ref{fig:crys_cden}.  VMC, DMC and extrapolated
results are shown for two different trial wave functions.  It can be seen that
the dependence of the extrapolated estimate on the trial wave function is much
smaller than for the raw VMC and DMC estimates, so we may have more confidence
in the extrapolated estimate of the charge density.


\section{Energy differences and energy derivatives}
\label{sec:energy differences and energy derivatives}

In electronic structure theory one is almost always interested in the
differences in energy between systems.  All electronic structure methods for
complex systems rely for their accuracy on the cancellation of errors in
energy differences.  In DMC this helps with all the sources of error mentioned
in section \ref{subsec:errors} except the statistical errors.  Fixed-node
errors tend to cancel because the DMC energy is an upper bound, but even
though DMC often retrieves 95\% or more of the correlation energy,
non-cancellation of nodal errors is the most important source of error in DMC
results.

\subsection{Energy differences in QMC}
\label{subsec:energy differences}

Correlated sampling methods allow the computation of the energy difference
between two similar systems with a smaller statistical error than those
obtained for the individual energies \cite{dewing_2002}.  Correlated sampling
is relatively straightforward in VMC, and a version of it is described in
section \ref{subsec:variance minimisation} in the context of optimising wave
functions by variance minimisation.


\subsection{Energy derivatives (forces) in QMC}
\label{subsec:forces}

Atomic forces are useful for relaxing the structures of molecules and solids,
calculating their vibrational properties, and for performing molecular
dynamics (MD) simulations.  It has proved difficult to develop accurate and
efficient methods for calculating atomic forces within QMC, although
considerable progress has been made in recent years.  Difficulties have arisen
in obtaining accurate expressions for DMC forces which can readily be
evaluated and in the statistical properties of the expressions, which are not
as advantageous as those for the energy.

According to the Hellmann-Feynman theorem (HFT), the derivative of the energy
with respect to a parameter $\lambda$ in the Hamiltonian is
\begin{eqnarray}
\label{eq:HFT force}
E^{\prime} & = & \frac{\int \Psi \, \hat{H}^{\prime} \, \Psi \, d{\bf R}}{\int
\Psi \, \Psi \, d{\bf R}} \;,
\end{eqnarray}
where the primes denote derivatives with respect to $\lambda$.  This
expression is valid when $\Psi$ is an exact eigenstate of $\hat{H}$.

Unfortunately the HFT is not normally applicable within QMC because the wave
functions are approximate.  Exact expressions for the VMC and DMC forces must
therefore contain additional Pulay terms which depend on $\Psi_{\rm
T}^{\prime}$.  To define the force properly it is therefore necessary to
define and evaluate $\Psi_{\rm T}^{\prime}$.

The DMC algorithm solves for the ground state of the fixed-node Hamiltonian
exactly and therefore the HFT holds.  Unfortunately the fixed-node Hamiltonian
is different from the physical Hamiltonian because it contains an additional
infinite potential barrier on the nodal surface of $\Psi_{\rm T}$ which forces
the DMC wave function $\phi_0$ to go to zero.  As $\lambda$ varies, the nodal
surface, and hence the infinite potential barrier, moves, giving a
contribution to $\hat{H}^{\prime}$
\cite{huang_2000,schautz_2000,badinski_2008a} which depends on $\Psi_{\rm T}$
and $\Psi_{\rm T}^{\prime}$ and is classified as a Pulay term.

The Pulay terms arising from the derivative of the mixed estimate of the
energy of equation (\ref{eq:diffusion_energy}) contain $\phi_0^{\prime}$, the
derivative of the DMC wave function.  This quantity cannot readily be
evaluated, and the approximation
\begin{eqnarray}
\label{eq:reynolds}
\frac{\phi_0^{\prime}}{\phi_0} & \simeq & \frac{\Psi_{\rm
T}^{\prime}}{\Psi_{\rm T}}
\end{eqnarray}
has normally been used
\cite{reynolds_1986b,assaraf_1999,casalegno_2003,assaraf_2003,lee_2005,badinski_2007,badinski_2008a,badinski_2008b,badinski_2008c}.
However, it leads to errors of first order in $(\Psi_{\rm T}-\phi_0)$ and
$(\Psi_{\rm T}^{\prime}-\phi_0^{\prime})$; therefore its accuracy depends
sensitively on the quality of $\Psi_{\rm T}$ and $\Psi_{\rm T}^{\prime}$, and
in practice this approximation is often inadequate.

The pure DMC energy,
\begin{eqnarray}
\label{eq:pure dmc energy}
E_{\rm D} = \frac{\int \phi_0 \hat{H} \phi_0 \, d{\bf R}}{\int \phi_0 \phi_0
\, d{\bf R}} \;,
\end{eqnarray}
is equal to the mixed DMC energy.  Forces may also be calculated within pure
DMC, and although this is more expensive it brings significant advantages.
The derivative $E_{\rm D}^{\prime}$ contains the derivative of the DMC wave
function, $\phi_0^{\prime}$.  However, Badinski \textit{et al.\@}
\cite{badinski_2008a} showed that $\phi_0^{\prime}$ can be eliminated from the
pure DMC expression, giving the exact result
\begin{eqnarray}
\label{eq:derivative of pure dmc energy 1}
E_{\rm D}^{\prime} & = & \frac{\int \phi_0 \phi_0 \, \phi_0^{-1}
\hat{H}^{\prime} \phi_0 \, d{\bf R}}{\int \phi_0 \phi_0 \, d{\bf R}} -
\frac{1}{2} \frac{\int \phi_0 \phi_0 \, \Psi_{\rm T}^{-2} |\nabla_{\bf R}
\Psi_{\rm T}| \Psi_{\rm T}^{\prime} \, d{\bf S}}{\int \phi_0 \phi_0 \, d{\bf
R}} \;,
\end{eqnarray}
where $d{\bf S}$ denotes an element of the nodal surface.  Unfortunately it is
not straightforward to evaluate integrals over the nodal surface.  The nodal
surface integral can be converted into a volume integral in which
$\phi_0^{\prime}$ does not appear using an approximation with an error of
order $(\Psi_{\rm T}-\phi_0)^2$, giving
\begin{eqnarray}
\label{eq:derivative of pure dmc energy 2}
E_{\rm D}^{\prime} & = & \frac{\int \phi_0 \phi_0 \, \left[\phi_0^{-1}
\hat{H}^{\prime} \phi_0 + \Psi_{\rm T}^{-1} \left(\hat{H}-E_{\rm D}\right)
\Psi_{\rm T}^{\prime} \right] \, d{\bf R}}{\int \phi_0 \phi_0 \, d{\bf R}} +
\\ && \frac{\int \Psi_{\rm T} \Psi_{\rm T} \, \left(E_{\rm L}-E_{\rm D}\right)
\Psi_{\rm T}^{-1} \Psi_{\rm T}^{\prime} \, d{\bf R}}{\int \Psi_{\rm T}
\Psi_{\rm T} \, d{\bf R}} + {{\cal{O}}[(\Psi_{\rm T}-\phi_0)^2}] \;.
\end{eqnarray}
This expression is readily calculable if one generates configurations
distributed according to the pure ($\phi_0^2$) and variational ($\Psi_{\rm
T}^2$) distributions.  The approximation is in the Pulay terms, which are
smaller in pure than in mixed DMC and, in addition, the approximation in
equation (\ref{eq:derivative of pure dmc energy 2}) is second order compared
with the first-order error in equation (\ref{eq:reynolds}).  Equation
(\ref{eq:derivative of pure dmc energy 2}) satisfies the zero variance
condition; if $\Psi_{\rm T}$ and $\Psi_{\rm T}^{\prime}$ are exact the
variance of the force obtained from equation (\ref{eq:derivative of pure dmc
energy 2}) is zero.  Equation (\ref{eq:derivative of pure dmc energy 2}) has
been used to obtain very accurate forces in small molecules
\cite{badinski_2008c,badinski_2009}.  The calculation of accurate DMC forces
is still in its infancy, but it does appear that equation (\ref{eq:derivative
of pure dmc energy 2}) offers a very promising way forward.

\section{Conclusions}
\label{sec:conclusions}

QMC methods provide a framework for computing the properties of correlated
quantum systems to high accuracy within polynomial time
\cite{note_exp_scaling}, facilitating applications to large systems.  They can
be applied to fermions and bosons with arbitrary inter-particle potentials and
external fields.  These intrinsically parallel methods are ideal for utilising
current and next-generation massively parallel computers.  Their accuracy,
generality and wide applicability suggest that they will play an important
role in improving our understanding of the behaviour of large assemblies of
quantum particles.

It is believed \cite{troyer_2005} that a complete solution to the fermion sign
problem may be impossible, and any exact fermion method may be exponentially
slow on a classical computer.  Accurate quantum chemistry techniques such as
the ``gold standard'' coupled cluster with single and double excitations and
perturbative triples [CCSD(T)] have been applied with considerable success to
correlated electron problems but, although they are also polynomial time
algorithms, their cost increases much more rapidly with system size than for
QMC methods.  DFT methods have proved extremely useful in describing
correlated electron systems, but there are many examples where the accuracy of
current density functionals has proved wanting.  It is important to remember
that trial wave functions for QMC calculations could be improved by developing
new wave function forms and better optimisation methods, whereas improving
approximate DFT methods requires the development of better density
functionals, which seems likely to be a much harder problem.

These considerations motivate the development of approximate QMC methods such
as those described in this review.  Although the basics of the DMC algorithm
used by Ceperley and Alder in 1980 \cite{ceperley_1980} have remained
unchanged, enormous progress has been made in using more complex trial wave
functions and in optimising the many parameters in them.  There is every
reason to believe that the current high rate of progress will continue for
many years to come.  Although these QMC methods will remain approximate, it is
clear that they can deliver highly accurate results provided the trial wave
functions are accurate enough.  Development of sophisticated computer packages
\cite{qmc_wiki} such as the \textsc{casino} code \cite{casino,casino_page}
should help to promote these methods.

\section{Acknowledgements}
\label{sec:acknowledgments}

We would like to thank all of our collaborators who have contributed so much
to our QMC project.  Much of this work has been supported by the Engineering
and Physical Sciences Research Council (EPSRC) of the UK\@.  NDD acknowledges
support from the Leverhulme Trust and Jesus College, Cambridge, and MDT
acknowledges support from the Royal Society. Computing resources were provided
by the Cambridge High Performance Computing Service.

\end{document}